%% file: main.tex
\documentclass[sigconf]{acmart}
\usepackage{hhline}
\usepackage{enumerate}
\usepackage{multirow}
\usepackage{array}
\usepackage{comment}
\usepackage{float}
\usepackage{graphicx}
\usepackage{tabularx}
\usepackage{makecell}
\usepackage{booktabs}
\usepackage{ulem}
\usepackage{lineno}
\usepackage{color}

\newcommand{\ken}[1]{\textcolor{black}{#1}}
\usepackage{colortbl} 
\usepackage{arydshln}
\usepackage{todonotes}
\usepackage{booktabs}
\usepackage{longtable}

\AtBeginDocument{%
  }

\copyrightyear{2026}
\acmYear{2026}
\setcopyright{cc}
\setcctype{by-nc-nd}
\acmConference[TEI '26]{Twentieth International Conference on Tangible, Embedded, and Embodied Interaction}{March 08--11, 2026}{Chicago, IL, USA}
\acmBooktitle{Twentieth International Conference on Tangible, Embedded, and Embodied Interaction (TEI '26), March 08--11, 2026, Chicago, IL, USA}
\acmPrice{}
\acmDOI{10.1145/3731459.3773325}
\acmISBN{979-8-4007-1868-7/2026/03}

\begin{document}

\input{SECTIONS/0_Title_Abstract}
\input{SECTIONS/1_Intro}

\input{SECTIONS/2_RW}

\input{SECTIONS/3_PD}
\input{SECTIONS/4_Study1}

\input{SECTIONS/5_Study2}

\input{SECTIONS/7_Discussion}

\input{SECTIONS/8_Conclusion}

\begin{acks}
This work was supported by JST BOOST, Japan, with the Grant Number JPMJBS2414.
\end{acks}

\bibliographystyle{ACM-Reference-Format}
\bibliography{reference}

\appendix

\pagebreak

\section{CORRESPONDENCE TABLE FOR EACH METHOD}\label{apendix-table}
Appendix A presents the correspondence tables used to assign instructional methods across participants in both user studies. In User Study 1, Table 4 shows how each participant was assigned audio descriptions, 2D tactile graphics, or 3D models for the six yoga poses, ensuring balanced exposure to all three conditions. Similarly, in User Study 2, Table 5 details the allocation of 2D tactile graphics and 3D models for the two sequences of calisthenic movements. These correspondence tables were designed to counterbalance the order of methods across participants and to minimize learning effects or bias in performance comparisons.

\begin{table}[htbp]
\centering
\caption{User Study 1: Correspondence Table for Each Method}
\Description{Table 4 shows which instructional method (Audio, 2D, or 3D) was assigned to each participant (P1–P10) for each of the six yoga poses, ensuring counterbalancing across Level 1 and Level 2 poses.}
\label{tab:user_study_1}
\setlength{\tabcolsep}{2pt}
\begin{tabular}{c|c c c|c c c}
\hline
& \multicolumn{3}{c|}{Level 1} & \multicolumn{3}{c}{Level 2} \\
\cline{2-7}
ID & (a) & (b) & (c) & (d) & (e) & (f) \\
 & Chair & \shortstack[c]{Crescent \\ Moon} & \shortstack[c]{Wind \\ Relieving} & Warrior I & Warrior II & \shortstack[c]{Reverse \\ Warrior} \\
\hline
P1 & Audio & 2D & 3D & 3D & Audio & 2D \\
P2 & 3D & Audio & 2D & Audio & 3D & 2D \\
P3 & 2D & 3D & Audio & 2D & 3D & Audio \\
P4 & Audio & 2D & 3D & 2D & Audio & 3D \\
P5 & 3D & Audio & 2D & 3D & 2D & Audio \\
P6 & 2D & 3D & Audio & 3D & 2D & Audio \\
P7 & Audio & 2D & 3D & Audio & 3D & 2D \\
P8 & 3D & Audio & 2D & 2D & Audio & 3D \\
P9 & 2D & 3D & Audio & Audio & 2D & 3D \\
P10 & Audio & 2D & 3D & 2D & 3D & Audio \\
\hline
\end{tabular}
\end{table}

\begin{table}[htbp]
\centering
\caption{User Study 2: Correspondence Table for Each Method}
\Description{Table 5 presents the instructional method assigned to each participant for Calisthenics 1 and 2. Each participant experienced one sequence with 2D and the other with 3D model, in counterbalanced order.}
\label{tab:user_study_2}
\begin{tabular}{c|c c}
\hline
ID & Calisthenics 1 & Calisthenics 2 \\
\hline
P1 & 2D & 3D \\
P2 & 3D & 2D \\
P3 & 2D & 3D \\
P4 & 3D & 2D \\
P5 & 2D & 3D \\
P6 & 3D & 2D \\
P7 & 2D & 3D \\
P8 & 3D & 2D \\
P9 & 2D & 3D \\
P10 & 3D & 2D \\
\hline
\end{tabular}
\end{table}

\pagebreak

\section{AUDIO DESCRIPTIONS FOR SIX YOGA POSES}\label{appendix-audio}
The following audio descriptions were originally developed in Japanese, the native language of the participants. For the purpose of this paper, they have been translated into English.

\paragraph{\textbf{(a) Chair Pose}}
This pose has three steps:
STEP 1: Stand with your feet together, balancing on your entire soles.
STEP 2: Bend both knees and lower your hips.
STEP 3: Raise both arms toward the ceiling.

\paragraph{\textbf{(b) Crescent Moon Pose}}
This pose has three steps: 
STEP 1: Stand with your feet close to each other, maintaining balance with both
soles. Feet are hip-width apart. 
STEP 2: Raise both arms from the sides upward, and bring your palms together in a prayer position.
STEP 3: Tilt your upper body to the right side.

\paragraph{\textbf{(c) Wind Relieving Pose}}
This pose has two steps:
STEP 1: Lie on your back on the floor.
STEP 2: Bend both knees and use your arms to hug them close to your chest.

\paragraph{\textbf{(d) Warrior I Pose}}
This pose has four steps:
STEP 1: Stand with your feet together, balancing on your entire soles.
STEP 2: Bend both knees and lower your hips.
STEP 3: Raise both arms toward the ceiling.
STEP 4: Step your left foot back significantly, turning your left toes 45 degrees outward while keeping your heel on the ground.

\paragraph{\textbf{(e) Warrior II Pose}}
This pose has four steps:
STEP 1: Step your feet apart to twice shoulder width, turning your right foot 90 degrees to the right and keep your left foot facing forward.
STEP 2: Bend your right knee to 90 degrees, aligning it over your heel.
STEP 3: Extend both arms to shoulder height.
STEP 4: Align your gaze with your right fingertips.

\paragraph{\textbf{(f) Reverse Warrior Pose}}
This pose has four steps:
STEP 1: Step your feet apart to twice shoulder width, with your right foot turned 90 degrees to the right and your left foot facing forward.
STEP 2: Bend your right knee to 90 degrees, aligning it over your heel.
STEP 3: Extend both arms to shoulder height, turn your right palm toward the ceiling, and stretch your right arm alongside your ear, leaning back.
STEP 4: Place your left hand on your left thigh.

\section{LIKERT SCALE RESULTS}

Appendix~C provides the detailed results of the 7‑point Likert scale questionnaires
administered in User Study~1 and User Study~2. These tables present individual
ratings for each participant (P1–P10) across the evaluation items and instructional
methods.

\begin{table*}[htbp]
  \centering
     \caption{User Study 1: Mean 7‑point Likert Scale Ratings for Three Instructional Methods}
     \Description{Table 6 displays individual 7-point Likert ratings from all participants (P1–P10) across six evaluation aspects for Audio, 2D, and 3D methods in yoga pose learning.}
  \begin{tabular}{|l|c|c|c|c|c|c|c|c|c|c|c|c|}
    \hline
    \textbf{Question} & \textbf{Method} & \textbf{P1} & \textbf{P2} & \textbf{P3} & \textbf{P4} & \textbf{P5} & \textbf{P6} & \textbf{P7} & \textbf{P8} & \textbf{P9} & \textbf{P10} & \textbf{Median} \\
    \hline
    \multirow{3}{*}{Q1: Enjoyment} & Audio & 5 & 7 & 5 & 7 & 5 & 7 & 3 & 7 & 4 & 1 & 5 \\
    \cline{2-13}
    & 2D & 6 & 7 & 2 & 5 & 4 & 4 & 4 & 7 & 7 & 5 & 5 \\
    \cline{2-13}
    & 3D & 6 & 7 & 7 & 6 & 7 & 7 & 6 & 7 & 3 & 7 & 7 \\
    \hline
    \multirow{3}{*}{Q2: Easiness} & Audio & 5 & 6 & 6 & 5 & 6 & 6 & 3 & 6 & 4 & 2 & 5.5 \\
    \cline{2-13}
    & 2D & 5 & 4 & 2 & 4 & 4 & 4 & 3 & 4 & 2 & 6 & 4 \\
    \cline{2-13}
    & 3D & 6 & 7 & 7 & 6 & 6 & 6 & 5 & 7 & 7 & 7 & 6.5 \\
    \hline
    \multirow{3}{*}{Q3: Effectiveness} & Audio & 5 & 7 & 6 & 5 & 5 & 7 & 2 & 5 & 5 & 3 & 5 \\
    \cline{2-13}
    & 2D & 5 & 7 & 2 & 5 & 4 & 4 & 3 & 2 & 3 & 7 & 4 \\
    \cline{2-13}
    & 3D & 6 & 7 & 7 & 6 & 6 & 6 & 6 & 7 & 7 & 7 & 6.5 \\
    \hline
    \multirow{3}{*}{Q4: Time Pressure} & Audio & 1 & 2 & 2 & 3 & 1 & 1 & 1 & 1 & 2 & 6 & 1.5 \\
    \cline{2-13}
    & 2D & 2 & 4 & 7 & 4 & 5 & 6 & 5 & 4 & 7 & 2 & 4.5 \\
    \cline{2-13}
    & 3D & 2 & 2 & 1 & 2 & 4 & 2 & 3 & 1 & 5 & 1 & 2 \\
    \hline
    \multirow{3}{*}{Q5: Motivation} & Audio & 4 & 6 & 4 & 5 & 6 & 2 & 4 & 6 & 6 & 2 & 4.5 \\
    \cline{2-13}
    & 2D & 4 & 5 & 1 & 4 & 4 & 2 & 3 & 4 & 2 & 6 & 4 \\
    \cline{2-13}
    & 3D & 4 & 7 & 7 & 7 & 7 & 2 & 6 & 7 & 6 & 7 & 7 \\
    \hline
    \multirow{3}{*}{Q6: Overall} & Audio & 6 & 7 & 6 & 6 & 6 & 5 & 6 & 5 & 7 & 5 & 6 \\
    \cline{2-13}
    & 2D & 4 & 5 & 1 & 5 & 4 & 1 & 5 & 4 & 3 & 6 & 4 \\
    \cline{2-13}
    & 3D & 5 & 7 & 7 & 7 & 6 & 4 & 7 & 7 & 5 & 7 & 7 \\
    \hline
  \end{tabular}
  \label{tab:us1_result}
\end{table*}

\begin{table*}[htbp]
  \centering
    \caption{User Study 2: Mean 7‑point Likert Scale Ratings for 2D and 3D Instructional Methods}
    \Description{Table 7 lists participant-wise Likert ratings comparing 2D and 3D methods for calisthenics learning across six aspects.}
  \begin{tabular}{|l|c|c|c|c|c|c|c|c|c|c|c|c|}
    \hline
    \textbf{Question} & \textbf{Method} & \textbf{P1} & \textbf{P2} & \textbf{P3} & \textbf{P4} & \textbf{P5} & \textbf{P6} & \textbf{P7} & \textbf{P8} & \textbf{P9} & \textbf{P10} & \textbf{Median} \\
    \hline
    \multirow{2}{*}{Q1: Enjoyment} & 2D & 5 & 6 & 2 & 4 & 6 & 1 & 1 & 7 & 4 & 6 & 4.5 \\
    \cline{2-13}
    & 3D & 6 & 6 & 7 & 6 & 5 & 1 & 7 & 7 & 6 & 7 & 6 \\
    \hline
    \multirow{2}{*}{Q2: Easiness} & 2D & 4 & 3 & 1 & 2 & 4 & 1 & 1 & 2 & 2 & 6 & 2 \\
    \cline{2-13}
    & 3D & 5 & 5 & 6 & 7 & 3 & 1 & 6 & 3 & 6 & 7 & 5.5 \\
    \hline
    \multirow{2}{*}{Q3: Effectiveness} & 2D & 5 & 7 & 2 & 4 & 6 & 1 & 5 & 5 & 4 & 6 & 5 \\
    \cline{2-13}
    & 3D & 6 & 7 & 7 & 7 & 5 & 1 & 6 & 7 & 6 & 7 & 6.5 \\
    \hline
    \multirow{2}{*}{Q4: Time Pressure} & 2D & 6 & 6 & 7 & 6 & 3 & 7 & 6 & 7 & 7 & 3 & 6 \\
    \cline{2-13}
    & 3D & 5 & 4 & 2 & 2 & 4 & 7 & 4 & 3 & 2 & 2 & 3.5 \\
    \hline
    \multirow{2}{*}{Q5: Motivation} & 2D & 5 & 4 & 2 & 4 & 6 & 1 & 1 & 5 & 4 & 6 & 4 \\
    \cline{2-13}
    & 3D & 5 & 4 & 6 & 7 & 5 & 1 & 6 & 6 & 6 & 7 & 6 \\
    \hline
    \multirow{2}{*}{Q6: Overall} & 2D & 5 & 4 & 2 & 5 & 5 & 1 & 3 & 6 & 4 & 6 & 4.5 \\
    \cline{2-13}
    & 3D & 6 & 5 & 7 & 7 & 4 & 4 & 6 & 7 & 6 & 7 & 6 \\
    \hline
  \end{tabular}
  \label{tab:us2_result}
\end{table*}

\end{document}

%% file: SECTIONS/0_Title_Abstract.tex
\title{Touching Movement:  3D Tactile Poses for Supporting Blind People in Learning Body Movements}


\author{Kengo Tanaka}
\affiliation{
  \institution{University of Tsukuba}
  \city{Tsukuba}
  \country{Japan}
}
\affiliation{
  \institution{Miraikan -- The National Museum of Emerging Science and Innovation}
  \city{Tokyo}
  \country{Japan}
}
\email{kengo.tanaka@digitalnature.slis.tsukuba.ac.jp}

\author{Xiyue Wang}
\affiliation{
  \institution{Miraikan -- The National Museum of Emerging Science and Innovation}
  \city{Tokyo}
  \country{Japan}
}
\email{wang.xiyue@lab.miraikan.jst.go.jp}
\author{Hironobu Takagi}
\affiliation{
  \institution{IBM Research - Tokyo}
  \city{Tokyo}
  \country{Japan}
}
\email{takagih@jp.ibm.com}

\author{Yoichi Ochiai}
\affiliation{
  \institution{University of Tsukuba}
  \city{Tsukuba}
  \country{Japan}
}
\email{wizard@slis.tsukuba.ac.jp}

\author{Chieko Asakawa}
\affiliation{
  \institution{Miraikan -- The National Museum of Emerging Science and Innovation}
  \city{Tokyo}
  \country{Japan}
}
\affiliation{
  \institution{IBM Research}
  \city{Yorktown}
  \country{USA}
}
\email{chiekoa@us.ibm.com}

\renewcommand{\shortauthors}{Tanaka et al.}


\begin{abstract}
Visual impairments create barriers to learning physical activities, since conventional training methods rely on visual demonstrations or often inadequate verbal descriptions. This research explores 3D-printed human body models to enhance movement comprehension for blind individuals. Through a participatory design approach in collaboration with a blind designer, we developed detailed 3D models representing various body movements and incorporated tactile reference elements to enhance spatial understanding. We conducted two user studies with 10 blind participants across different activities: static yoga poses and sequential calisthenic movements. The results demonstrated that 3D models significantly improved understanding speed, reduced questions for clarification, and enhanced movement accuracy compared to conventional teaching methods. Participants consistently rated 3D models higher for ease of understanding, effectiveness, and motivation.

\end{abstract}

\ccsdesc[500]{Human-centered computing~Accessibility design and evaluation methods}
\ccsdesc[500]{Human-centered computing~Accessibility technologies}



\keywords{Tactile graphics, 3D model, Accessibility, Visual impairment}


\begin{teaserfigure}
  \includegraphics[width=\textwidth]{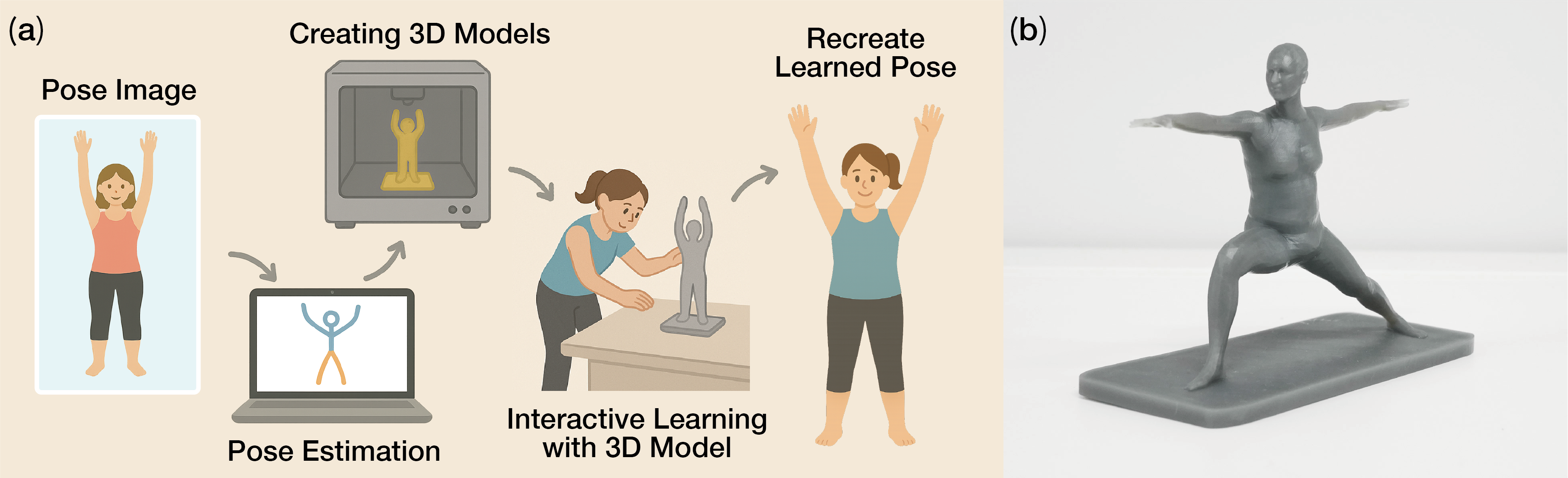}
  \caption{(a) Proposed system workflow showing the process from an initial pose image to pose recreation. The workflow includes pose estimation from 2D images, generating 3D models based on the estimated poses, and fabricating them with a 3D printer. Users first learn by tactually exploring the printed models and then reproduce the learned poses.
(b) Example of a 3D-printed model representing a yoga pose used in the user study.}
  \Description{Figure 1 illustrates the overall workflow of the proposed system for pose learning using 3D-printed tactile models.
(a) The figure shows four sequential steps from left to right:
First, an illustration of a woman with arms raised represents the pose image input.
Second, the pose estimation step is represented by a laptop screen displaying a simplified skeleton overlay derived from the image.
Third, the 3D model creation step is depicted as a 3D printer fabricating a yellow model in a raised-arm pose.
Fourth, in the interactive learning step, a user is shown touching and examining the 3D-printed model to understand the pose before attempting to physically recreate it. Arrows indicate the direction of the workflow, culminating in the user replicating the learned pose.
(b) A color photograph shows a 3D-printed model representing a yoga pose. The model depicts a human figure in a wide-legged stance with arms extended horizontally, standing on a rectangular base designed for tactile stability.}
  \label{fig:teaser}
\end{teaserfigure}

\maketitle

%% file: SECTIONS/1_Intro.tex
\section{INTRODUCTION}
Participation in regular physical activities such as sports, exercise, and dance is reported to contribute to maintaining and enhancing health and quality of life (QoL) ~\cite{Windle01082010, Wilcox2003, inbook}.
However, learning these activities typically relies on visually oriented demonstrations, where learners are expected to observe and mimic movements shown by the instructors~\cite{Duggar01051968}, which pose a fundamental accessibility challenge for people who are visually impaired.
To address this gap, verbal explanations are added to describe physical movements ~\cite{seham2012dance, Haegele2019}.
However, such descriptions frequently suffer from inconsistent terminology across instructors, leading to ambiguity and confusion ~\cite{seham2012dance, 02646196211059756}. Moreover, verbal instruction alone struggles to convey subtle aspects of movement, such as timing, posture, and style ~\cite{DeSilva2023}.
Another common approach is direct physical guidance by instructors, where movements are demonstrated through touch ~\cite{Megan2006, seham2012dance}. While this method can offer some additional clarity, it also raises concerns regarding personal boundaries, privacy, and psychological comfort ~\cite{Holloway2022, Megan2006}.

In response to these challenges, accessibility research in physical education for individuals with visual impairments has progressed. A number of studies have focused on enabling blind individuals to participate in physical activities such as sports and exercise ~\cite{Strobel2025}. Specifically, researchers have explored ways to convey human motion using exercise that integrate voice guidance with posture recognition ~\cite{Dias2019, RectorYoga2014, RectorYoga2013, Rector2017}, as well as through refreshable tactile displays (RTDs) ~\cite{Holloway2022}. However, these methods still pose challenges in effectively conveying complete human movement patterns due to several limitations: voice guidance suffers from variability in verbal nuance interpretation ~\cite{DeSilva2023}; RTDs have limited resolution and screen size; and RTDs also struggle with representing multiple body parts moving simultaneously ~\cite{Holloway2022}.
Furthermore, while studies exist on the utilization of three-dimensional models for instructing fundamental postures of traditional dance ~\cite{Sucharitakul19122023}, discourse concerning bodily movements remains limited in scope. Research on three-dimensional representations for understanding movement through tactile sensation is insufficient, and comprehensive verification of effective methodologies for communicating complex physical movements through three-dimensional models has yet to be adequately conducted.

To address this critical gap, this research explores the potential of using 3D-printed human body models to support tactile learning and non-visual understanding of physical movements. In this paper, we present an approach that helps blind individuals learn physical movements through hands-on exploration of detailed physical models.
We created such models by first generating a 3D models using automated 3D pose reconstructing tools ~\cite{goel2023humans4dreconstructingtracking} and then printing the models with a 3D printer. 
Our development process followed an iterative design approach, where we collaborated closely with a blind participant designer (P0). Across multiple design cycles, P0’s tactile evaluations and feedback guided improvements in clarity, usability, and communicative effectiveness.

To evaluate whether these tactile representations support more accurate understanding of body posture and movement than traditional methods, we conducted two progressive user studies with 10 blind participants, covering a broad spectrum of body poses and movements, from static to dynamic: Study 1 evaluated static yoga poses, comparing 3D models against verbal instructions and 2D tactile graphics; Study 2 examined sequential calisthenics exercise movements using 3D models and 2D tactile graphics.
Results across all studies demonstrated that 3D models improved understanding speed, reduced questions needed for comprehension, and enhanced movement reproduction accuracy. Participants consistently rated 3D models higher for ease of understanding, effectiveness, and motivation compared to alternative methods. 

The main contributions of this study are as follows:
\begin{itemize}
    \item A participatory investigation of the design of 3D-printed physical models to support blind individuals in understanding and reproducing body postures and movements.
    \item An empirical comparison of 3D models with conventional teaching methods (verbal instructions and 2D tactile graphics), and demonstrating better objective performance and higher subjective learner ratings.
\end{itemize}

This work represents an important step toward developing more inclusive instructional methods that enhance independence and participation in physical activities for blind individuals.
By demonstrating the potential of tangible 3D representations, this study not only addresses gaps in non-visual movement instruction but also lays the groundwork for future research exploring tactile, multisensory, and embodied approaches to accessible physical education.

%% file: SECTIONS/2_RW.tex
\section{RELATED WORK}

\subsection{Access to and Challenges in Physical Activity for Blind Individuals}
Blind individuals rely on their tactile and auditory senses to understand the world around them; however, comprehending movement through these senses is challenging. As Holloway et al. ~\cite{Holloway2022} noted, blind individuals face unique challenges when learning bodily movements because they cannot apply the common ``see and imitate'' learning process.
Research indicates that participation rates in physical activity among blind individuals are lower than for sighted individuals, and this difference impacts their health status and quality of life ~\cite{Michele2007, Weil2002, Ackley2009, Seham2015}. Primary barriers to accessing physical activity programs include mobility difficulties ~\cite{Jaarsma2014, Lee2014}, safety concerns ~\cite{Audun2017, India2021}, and the absence of inclusive educational programs ~\cite{Boswell2022}. These factors affect the physical condition of blind individuals ~\cite{Michele2007, Haegele2019} and lead to reduced opportunities for social interaction ~\cite{Lieberman2002, Seham2015}.
The principal non-visual technique for teaching bodily movements to blind individuals is verbal instruction. As Duggar ~\cite{Duggar01051968} pointed out, education in physical movement typically relies on visual explanations. However, Seham \& Yeo ~\cite{Seham2015} emphasized that when teaching blind individuals, the initial concept of each step, posture, position, and movement must be conveyed without relying on visual cues.
To supplement verbal instructions, physical interactions through touch and demonstrations are commonly employed ~\cite{Megan2006, seham2012dance}. Instructors may physically position parts of the learner's body to convey rhythm or movement, or learners may touch the demonstrator to understand movements. Each of these methods has inherent advantages and challenges. As drawbacks, for example, these approaches can sometimes lead to misunderstandings, may not be suitable for students sensitive to touch, and may invade the personal space of instructors or trainees ~\cite{Holloway2022, Megan2006}. Additionally, De Silva et al. ~\cite{DeSilva2023} have identified the challenges and needs in teaching physical movements to the visually impaired and have proposed potential areas for solutions.

\subsection{Assistive Technologies for Motion Perception and Learning}
While there are numerous guidelines for designing tactile graphics ~\cite{BANA2010, RoundTable2005}, there is a lack of advice on conveying physical movement and change. Although tactile graphics technology traditionally focuses on static representations, recent developments have introduced various dynamic approaches.

For individuals with visual impairments, Refreshable Tactile Displays (RTDs) offer a tactile approach to expressing movements of objects and the body. These displays use pins of varying heights to create tactile animations.
RTDs have been employed to convey scenarios in sports such as soccer ~\cite{Ohshima2021}, and floor volleyball ~\cite{Kobayashi2021}. Additionally, Holloway et al. ~\cite{Holloway2022} have used RTDs to represent human movement through tactile animations. However, these displays face limitations in terms of resolution and screen size, making it challenging to track the movements of multiple limbs.

In the domain of auditory feedback technologies, research has been conducted to convey the movement of external objects through sound in sports such as hockey ~\cite{Hockey2022, Doherty2022}, goalball ~\cite{Miura2018Goalball, WatanabeMMSO22}, sound table tennis ~\cite{Miura2018TableTennis}, badminton ~\cite{Sadasue2021}, and tennis ~\cite{Jain2023}. Additionally, there are auditory-based solutioDiasns focusing on spatial awareness concerning the orientation of the entire body in activities such as swimming ~\cite{Swim2017, Oommen2018}, and skiing ~\cite{Miura2023}. Other studies focus on the orientation of specific body parts in activities such as rock climbing ~\cite{Ramsay2020}, yoga ~\cite{Rector2017, RectorYoga2014, RectorYoga2013}, and dance ~\cite{Dias2019, Katan2016}.
Dias et al. ~\cite{Dias2019} developed a platform that enables visually impaired students to learn dance remotely using motion tracking and auditory feedback. Rector et al. ~\cite{Rector2017} created a yoga system employing pose estimation and auditory feedback to allow visually impaired individuals to exercise safely. However, limitations have been indicated in using audio to convey nuances of direction and specific body movements for spatial awareness.

Research has also explored multimodal approaches combining tactile and auditory feedback. Tactile feedback has been used as a mechanism in training body movements for blind and visually impaired (BLV) individuals in activities such as running ~\cite{Júnior2021, Rector2018}, skiing ~\cite{Skier2016}, exergame ~\cite{VI-Tennis2010, Vi-bowling2010}, yoga ~\cite{Islam2022}, and dance ~\cite{GuiDance2018}.
For instance, in yoga, Islam et al. ~\cite{Islam2022} tested the efficacy of vibrational and auditory feedback for visually impaired individuals, demonstrating that vibrational feedback was effective and time-efficient in correcting yoga poses.
In dance, Camarillo-Abad et al. ~\cite{GuiDance2018} showed that it is possible to guide individuals in remote locations using only vibrotactile feedback. De Silva et al. ~\cite{de_silva_sensing_2025} conducted dance workshops using sensory probes, enhancing the accessibility of dance.
While multimodal feedback, such as tactile and auditory, is effective in conveying directional instructions and timing for body parts in physical movement learning, it has been limited to yoga and dance studies.

\ken{\subsection{3D Model-Based Representations for Spatial and Movement Understanding}}

\ken{Recent research on 3D-printed and interactive tactile models has expanded access to spatial learning for BLV users. Holloway et al.~\cite{Holloway2018} showed that 3D-printed maps improve recall and understanding of height and layout compared to traditional tactile graphics. Wang et al. further developed interactive 3D systems such as BentoMuseum and TouchPilot, which incorporate audio guidance to support multilevel navigation and systematic exploration of complex 3D structures~\cite{BentoMuseum2022, TouchPilot2023}. Reinders et al.~\cite{Reinders2023DIS, Reinders2025} introduced multimodal Interactive 3D Models (I3Ms) combining tactile, auditory, and conversational feedback to enhance engagement and agency during exploration, while Tsutsui et al.~\cite{Tsutsui2025ASSETS} conducted design research examining how BLV users and tactile educators collaboratively use multiple 3D models for science communication. 
These studies demonstrated the effectiveness of 3D-printed models in supporting non-visual understanding, as well as the potential of their interactive and multimodal augmentations.}

\ken{From a cognitive perspective, 3D tactile models make spatial structure directly perceivable, reducing the mental effort required to interpret spatial relationships compared to 2D tactile graphics~\cite{Holloway2018}. According to Cognitive Load Theory~\cite{Morrison2005, Paas_Sweller_2014}, effective learning materials should reduce extraneous cognitive load and align perceptual input with human cognitive capacities. Empirical studies have shown that multimodal presentations combining touch and sound can lower cognitive load, increase engagement, and improve task performance for blind learners~\cite{Mackowski2022, Mackowski2023}. These findings suggest that 3D tactile models, by allowing direct perception of spatial relationships such as depth and proportion, can naturally distribute cognitive demands across sensory modalities, supporting more efficient comprehension and retention of complex spatial information.}

\ken{Extending these insights to human movement, educators have used manipulable anatomical models to convey three-dimensional body configurations~\cite{DeSilva2023}. Sakashita et al.~\cite{Sakashita2017} generated 3D anatomical forms from dance performances to document choreography, but their system targeted sighted users and did not assess accessibility for BLV learners. Sucharitakul et al.~\cite{Sucharitakul19122023} employed 3D-printed tactile models to teach basic Thai dance poses, yet the evaluation focused on a limited set of movements and offered little discussion of how visually impaired adults perceive and interpret 3D objects. While these studies show that 3D models aid understanding of certain postures, comprehensive investigations of their role in movement learning remain scarce. Building on this gap, we propose a 3D model–based system for BLV learners that supports stepwise understanding of multi-pose movements, addressing challenges such as simultaneous limb actions and pose continuity through stakeholder-informed design and evaluation.}

%% file: SECTIONS/3_PD.tex
\section{PARTICIPATORY DESIGN OF 3D MODELS}
We employed a participatory design approach, collaborating with a blind designer and researcher (P0, female, 66 years old) through three 60-minute sessions to develop both 3D models that support tactile learning of physical movements and a comprehensive workflow for their creation. \ken{P0 has been blind since the age of 14 and possesses extensive experience in teaching and evaluating tactile materials such as braille and physical models. This long-term engagement with tactile design enabled P0 to offer informed feedback on model detail, orientation cues, and usability.} Through iterative co-design sessions, we established a user-centered development process that ensures the resulting models meet the specific needs and preferences of visually impaired learners. All feedback from P0 was collected through audio recordings and detailed field notes, ensuring traceability and accuracy of the input. The design process followed three iterative steps: (1) generating 3D poses from pose estimation data and exploring their potential applications for tactile representation, (2) creating tactile content through collaborative refinement sessions to determine optimal physical characteristics and informational elements, and (3) designing tactile support structures and interaction mechanisms for optimal haptic exploration and learning effectiveness.

\subsection{Session 1: Initial Concept Development and Problem Recognition}
\subsubsection{Problem Identification and Initial Prototyping}
Our first session began with P0 sharing experiences as a blind interaction designer and researcher. P0 had previously participated in online yoga lessons but found it difficult to follow along with audio-only instructions and struggled to understand the physical movements. P0 highlighted challenges in learning movements through verbal instructions in an online yoga program, stating: \textit{``Even with instructions like 'Keep your back straight and extend your arms to the sides,' the expected movement was hard to visualize and required trial and error.''} This shows the struggle visually impaired individuals face in movement-based activities, highlighting the need for tactile verification to convey posture nuances, motivating the development of 3D models.

\subsubsection{Creation and Evaluation of Initial Prototypes}
Motivated by this insight, we created small-scale 3D models depicting various human postures using ``Humans in 4D''~\cite{goel2023humans4dreconstructingtracking}, which accurately represents subtle human movements and posture changes from a single-view RGB video. While pose estimation has primarily been used for applications such as motion analysis and real-time feedback in sports and rehabilitation, our approach uniquely leverages it to produce 3D-printed tactile models, highlighting its potential as educational materials for non-visual learning. Rather than using conventional approaches with models featuring movable parts \cite{DeSilva2023}, we chose this pose estimation technology to explore the potential of applying human pose estimation to tactile learning. This approach enables systematic generation of any specific posture required for learning. This digital approach creates reusable assets that can be instantly shared, scaled to different sizes for various learning contexts, and systematically archived for educational use. 

P0 evaluated these prototypes through tactile exploration, providing comprehensive feedback on shape recognition. P0 noted several key aspects that facilitate spatial understanding: \textit{``The curvature of the back and nose protrusion clarify orientation, which is essential for spatial recognition.''} P0 further explained that \textit{``checking knee bends helps to understand posture direction and limb relationships,''} while also emphasizing the importance of detailed features: \textit{``I can understand the direction of palms and toes, which are crucial for replicating the exact posture.''} Additionally, P0 highlighted how \textit{``the weight distribution and balance of the pose become clear through touching the model's stance.''} This comprehensive feedback underscores the intuitive understanding provided by tactile examination of 3D models generated from real human pose data.

\subsection{Session 2: Content Creation and Target Selection}
Building on the findings from Session 1, and following the same procedure for data collection, we refined the design focus with our blind co-designer (P0) through a 60-minute follow-up session.

Based on P0’s evaluation of initial prototypes, we confirmed three key directions:

\begin{itemize}
\item Target both static poses and dynamic movements.
\item Determine optimal model size and level of detail for tactile exploration.
\item Focus on movements that are typically difficult to convey verbally, especially those involving simultaneous limb coordination.
\end{itemize}

\subsubsection{Selection of Target Poses and Movements}
Through consultation with P0, we focused on two areas: yoga and calisthenics. P0 emphasized that \textit{``These are activities that visually impaired individuals wish to participate in but have found difficult due to the lack of appropriate learning methods.''} Research by De Silva et al. \cite{DeSilva2023} and Holloway et al. \cite{Holloway2022} confirmed a strong interest among BLV individuals in learning activities like yoga, dance, and martial arts, thus aligns with our selection focus.

\begin{figure*}
  \centering
  \includegraphics[width=\textwidth]{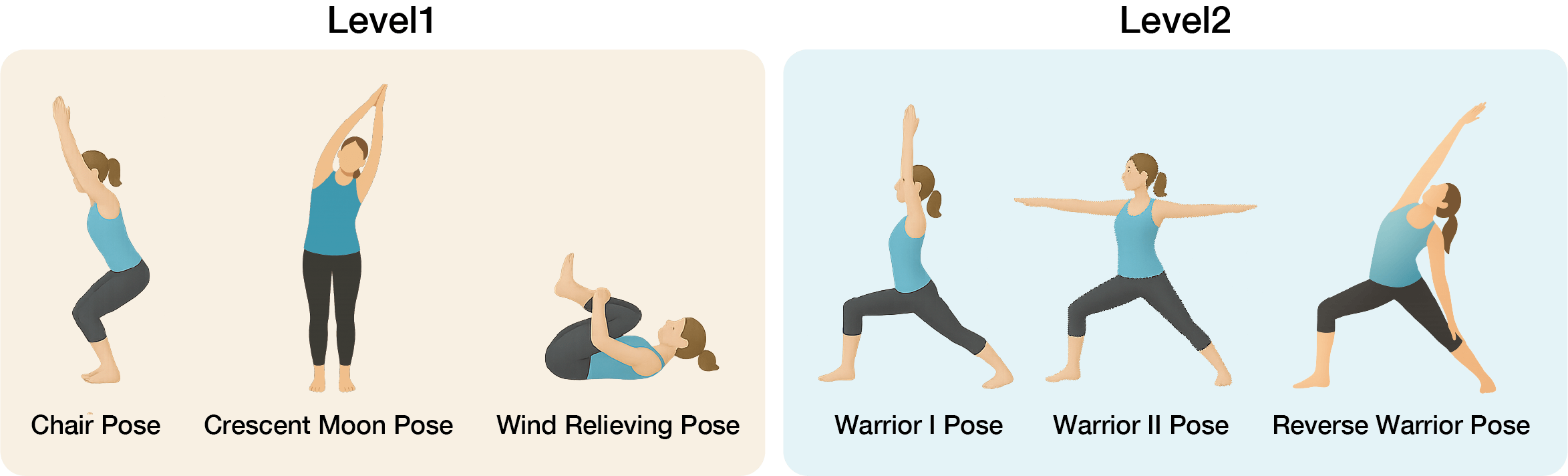}
  \caption{Six yoga poses categorized by physical difficulty level. Level 1 (left) shows three beginner poses: Chair Pose, Crescent Moon Pose, and Wind Relieving Pose. Level 2 (right) presents three intermediate poses: Warrior I Pose, Warrior II Pose, and Reverse Warrior Pose. }
  \Description{Figure 2 shows six yoga poses grouped by difficulty level. The left side, labeled “Level 1” with a beige background, presents three beginner poses: Chair Pose, Crescent Moon Pose, and Wind Relieving Pose. The right side, labeled “Level 2" with a light blue background, displays three intermediate poses: Warrior I Pose, Warrior II Pose, and Reverse Warrior Pose. }
  \label{fig:yoga}
\end{figure*}

\begin{figure*}
  \centering
  \includegraphics[width=\textwidth]{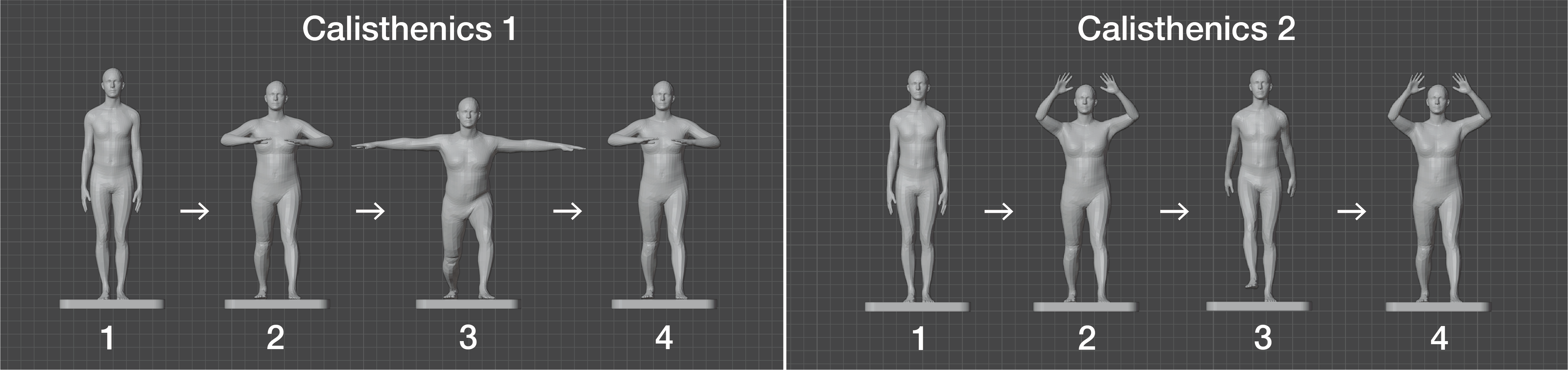}
  \caption{Calisthenics Poses showing two different sequences of four progressive movements.}
  \Description{Figure 3 displays two calisthenics sequences, each consisting of four progressive poses. On the left side, labeled “Calisthenics 1,” four numbered 3D human models are shown in a horizontal sequence, with rightward arrows indicating the progression. On the right side, labeled “Calisthenics 2,” another set of four numbered 3D models is presented in a similar layout. Both sequences are designed to depict step-by-step movements used in the study.}
  \label{fig:taiso}
\end{figure*}

\subsubsection{Yoga Pose Selection}
We collaborated with two certified yoga instructors (I1: a 23-year-old female \ken{with 2 years of teaching experience,} and I2: an 83 year-old male \ken{with 60 years of teaching experience} ), of whom I2 had prior experience teaching students with visual impairments. Together with P0, we selected six beginner-level poses, grouped into two tiers (Level 1 and Level 2) as shown in Figure~\ref{fig:yoga}. The selection criteria were: (1) high prevalence in beginner yoga classes, (2) clear contrasts in limb positioning, and (3) feasibility for tactile representation without excessive complexity. In addition, we intentionally included a floor-based pose (Wind Relieving) that involves lying on a mat. Prior research such as Eyes-Free Yoga \cite{Rector2017, RectorYoga2014, RectorYoga2013} has shown limitations in recognizing seated or lying poses with Kinect-based systems. By incorporating a lying-down posture, we aimed to address this gap while also exploring how tactile 3D models could convey depth cues that 2D tactile graphics often fail to represent. P0 confirmed that including this type of posture was essential for reflecting the real difficulties faced by beginners in yoga practice.

P0 noted that these poses reflected common challenges faced in non-visual learning, especially when overlapping limbs or subtle joint rotations were involved. Level 1 poses (Chair, Crescent Moon, Wind Relieving) emphasized stability and straightforward alignment, serving as an accessible foundation. Level 2 poses (Warrior I, Warrior II, Reverse Warrior) introduced wider stances, increased complexity of body alignment, and balance elements, providing a progressive challenge.

\subsubsection{Calisthenic Movement Selection}

Inspired by Japan’s popular ``Radio Calisthenics'' and ``Minna no Taiso (Calisthenics for Everyone)''\footnote{\url{https://www.jp-life.japanpost.jp/health/radio/}}, we designed two movement sequences, each consisting of four progressive poses (Figure~\ref{fig:taiso}). The number of poses was set at four based on P0’s feedback that longer sequences increased memory burden and risked confusion, while fewer than four provided insufficient flow to evaluate continuity. Both sequences began and ended in a basic standing position. The second and fourth poses were intentionally repeated, allowing us to assess learners’ ability to track circularity and continuity in movement. P0 highlighted that the inclusion of repeated poses would help blind learners verify orientation consistency and recognize rhythm. 

The selected calisthenic movements involved coordinated arm and leg actions (forward, backward, upward, sideways), aligning with P0’s observation that \textit{``simultaneous movements of arms and legs are the most confusing to follow without tactile references.''} While it was not clear whether such complex, multi-limb movements could be intuitively understood through tactile 3D models, we intentionally included them to investigate this question. This choice also reflects prior findings by Holloway et al.~\cite{Holloway2022}, who reported that when conveying Tai Chi movements using a refreshable tactile display, participants struggled to track simultaneous limb actions due to resolution and screen size limitations. By testing whether 3D-printed models can better support the comprehension of these challenging movements, we aimed to evaluate their potential as an effective tactile learning medium.

\subsubsection{3D Model Generation and Size Determination}
This session highlighted both the potential of a semi-automated pipeline for creating tactile learning materials and the determination of an appropriate model size for tactile exploration.

By leveraging pose estimation technologies such as Humans in 4D, we were able to generate accurate 3D body postures directly from single-view video frames, significantly reducing the manual effort traditionally required to design each model. While manual refinements in Blender were still necessary for subtle joint orientations and posture adjustments, P0 acknowledged that \textit{``having the basic pose automatically generated made the models more consistent and easier to refine.''} This semi-automated process demonstrates how future systems could streamline the creation of a wide variety of educational models, lowering the production barrier and enabling rapid adaptation to learners’ needs.

To evaluate tactile manageability, we fabricated Warrior II models at five different scales, with heights of 7.5 cm, 8.5 cm, 10.5 cm, 12.5 cm, and 15 cm from the ground to the top of the head. P0 compared these versions and concluded that the 10.5 cm height provided the best balance: it was large enough to perceive fine details such as finger orientation and joint angles, while still compact enough to be comfortably held in both hands. Based on this feedback, we standardized the scaling ratio of all models so that a standing Warrior II pose measured 10.5 cm, and applied this ratio consistently across all yoga and calisthenic poses in the user studies to reduce cognitive load and support comparability between activities.

\begin{figure*}
  \centering
  \includegraphics[width=\textwidth]{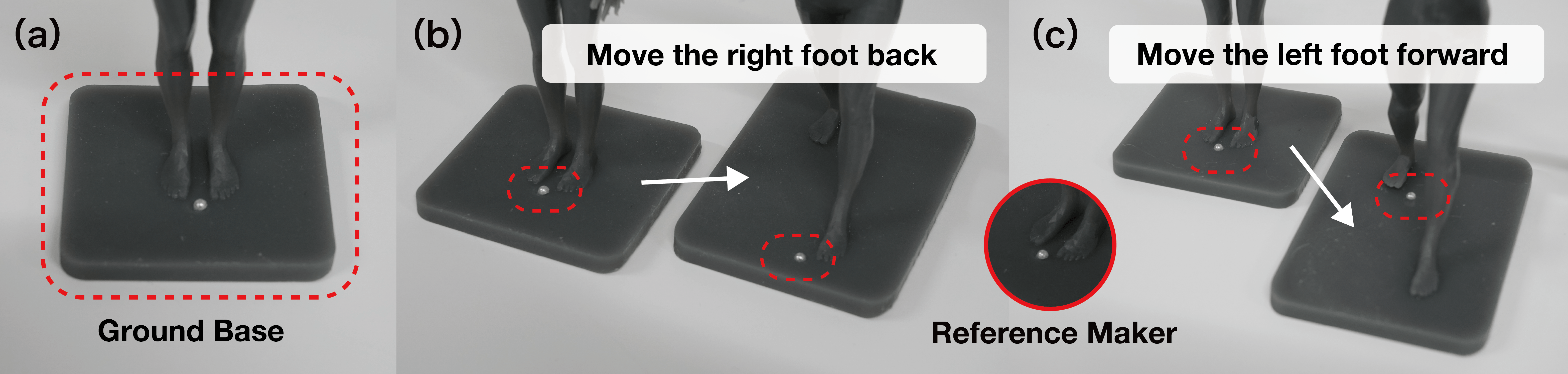}
  \caption{Base design components of the proposed 3D modeling system. (a) A ground base with a reference marker, indicating floor orientation and helping users sense weight distribution and body balance. (b) Instruction to move the right foot backward using the reference marker. (c) Instruction to move the left foot forward using the reference marker.}
  
  \Description{Figure 4 illustrates the base design components of the 3D modeling system, focusing on the role of reference markers. (a) Shows a square ground base with a small tactile reference marker embedded near the front center. This marker provides an absolute starting position for foot placement. (b) Demonstrates an instruction to move the right foot backward relative to this reference point, indicated by an arrow and highlighted marker. (c) Similarly, illustrates moving the left foot forward from the same fixed reference point.}
  \label{fig:base}
\end{figure*}

\subsection{Session 3: Support Structure Design and Refinement}

In the final co-design session, we focused on creating support structures that enhanced stability and tactile clarity of the 3D models, ensuring that participants could concentrate on learning the body postures without distraction. This included decisions on model bases, orientation markers, and material selection. 

\subsection{Base Design Development}
P0 emphasized that \textit{``clearly defining where the body meets the ground makes it easier to grasp the pose and understand balance.''} To address this, we introduced a ground base that ensured stability and provided clear contact points with the floor (Figure~\ref{fig:base} (a)). The base minimized wobbling during tactile exploration and offered a consistent frame of reference for distinguishing standing versus lying poses. We selected a square base for static poses and a slightly elongated rectangular base for calisthenic movements, ensuring both stability and tactile clarity.

\subsection{Reference Markers for Orientation}
In dynamic movements such as calisthenics, accurate recognition of which foot moves during a sequence was identified as a recurring challenge. To address this, we placed a small tactile reference marker on the base between the feet in the initial standing position (Figure~\ref{fig:base} (b), (c)). This marker served as a stable point of reference, allowing participants to track which foot remained in place and which moved in subsequent poses. For example, when the next pose required stepping forward with the right foot, the marker remained adjacent to the left foot, indicating that the left foot stayed fixed while the right foot advanced. P0 noted that \textit{``the marker made it clear which foot had moved, so I could immediately understand the change of position.''} By providing this simple but consistent cue, the marker supported spatial continuity without the need for rhythm or temporal information.

\begin{figure*}
  \centering
  \includegraphics[width=\textwidth]{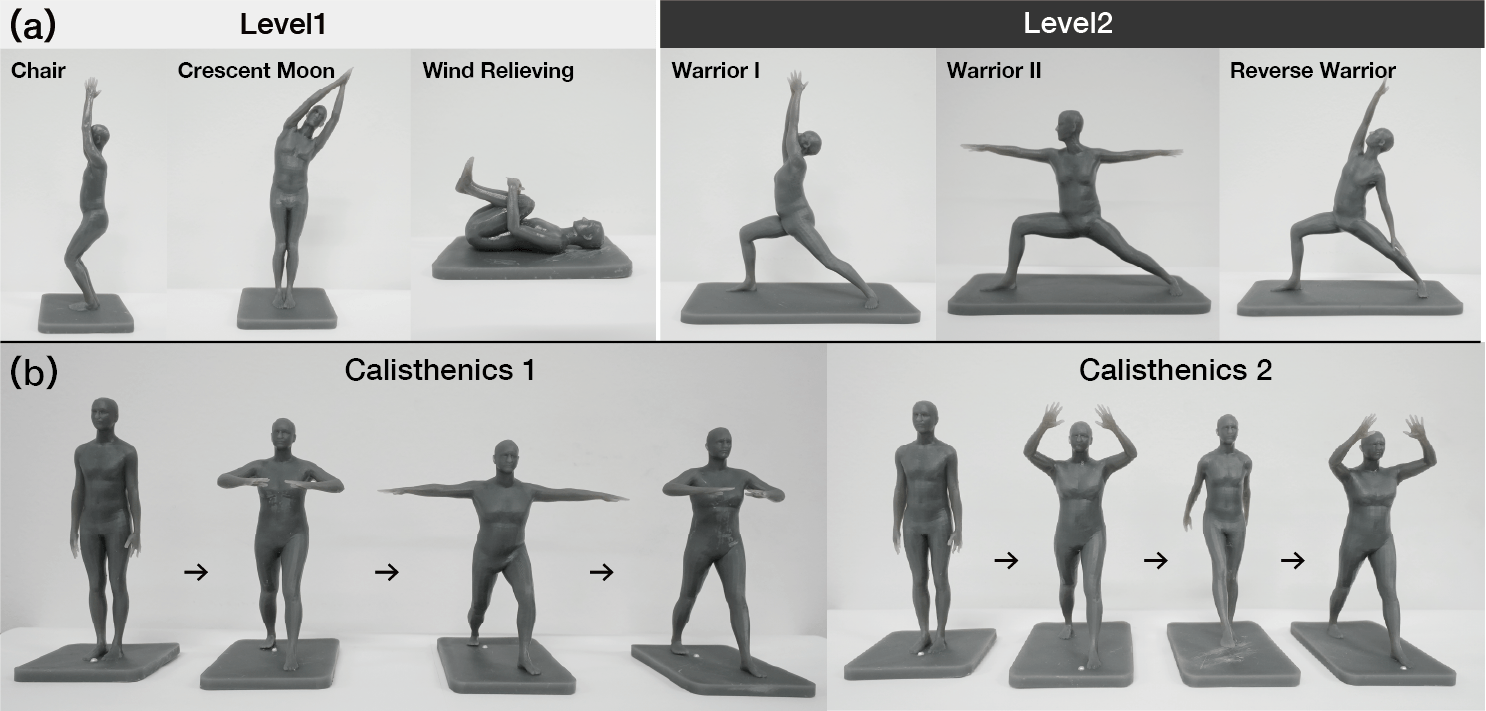}
    \caption{
    Final design of the 3D tactile pose models used in the study. 
    All figures in (a) and (b) are mounted on flat bases to enhance tactile stability and alignment. 
    (a) Six yoga poses divided into two difficulty levels: \textit{Level 1} includes Chair, Crescent Moon, and Wind Relieving poses; \textit{Level 2} includes Warrior I, Warrior II, and Reverse Warrior poses. 
    (b) Two calisthenic movements, \textit{Calisthenics 1} and \textit{Calisthenics 2}, each represented as a four-step sequence using distinct 3D-printed figures to show transitional postures. 
    Reference markers on the bases are included to help users perceive foot position and forward-backward motion.
    }
  \Description{Figure 5 presents the finalized 3D-printed tactile models used in the study. (a) Shows six yoga poses divided into two difficulty levels, each figure mounted on a flat base for tactile alignment: Level 1 includes Chair, Crescent Moon, and Wind Relieving poses; Level 2 includes Warrior I, Warrior II, and Reverse Warrior poses. (b) Displays two calisthenic exercises, Calisthenics 1 and Calisthenics 2, each represented as a sequence of four 3D models arranged left to right with arrows indicating progression. All bases include reference markers to help users understand foot placement and forward-backward transitions.}
  \label{fig:final}
\end{figure*}

\subsection{Printing Materials and Fabrication}
The finalized models were fabricated using a Formlabs Form 3L 3D printer. After printing, all residual support structures were carefully removed, and the surfaces were sanded to eliminate rough edges. This finishing process ensured that the models could be comfortably explored by touch, preventing any unwanted tactile noise from interfering with the perception of body posture. To ensure safety and consistency, we applied the same finishing process to all models and confirmed with P0 that the surfaces felt smooth and did not cause discomfort during extended tactile exploration. \ken{Although the SLA resin used in fabrication was more costly than thermoplastics such as PLA, it provided sufficient strength and durability for study-scale use without noticeable wear or damage. This choice ensured smooth surface quality and consistent tactile clarity throughout repeated exploration.}

\subsection{Summary of Refinements}
Through this iterative process, we refined the models to balance durability, tactile clarity, and orientation cues. These refinements, co-developed with P0, ensured that the models could be explored comfortably, clearly communicated posture and limb orientation, and remained stable throughout repeated use. The finalized designs were then employed in the subsequent user studies to evaluate their effectiveness in supporting non-visual comprehension of body movements.

%% file: SECTIONS/4_Study1.tex
\section{USER STUDY 1: Comparing 3D Models With 2D Tactile Graphics and Audio Instruction for Comprehending Yoga Poses}

\subsection{Apparatus for Comparison}

To investigate the effectiveness of 3D models in conveying static yoga poses, we created the apparatus of traditional methods, audio instructions, and 2D tactile graphics.

\subsubsection{Audio Design}
We developed step-by-step verbal instruction scripts for the yoga poses. The first draft was created by the authors based on publicly available step-by-step yoga guides intended for general audiences. This draft was then refined through two rounds of review by two certified yoga instructors (I1 \& I2). Their feedback ensured that the scripts maintained clarity, conciseness, and logical structure without relying on visual cues. Following a step-by-step structure was also supported by prior work; Mohanty et al.~\cite{Mohanty2016} demonstrated that using sequential, step-based instructions was effective in teaching yoga to children with visual impairments. The final version of each script presented the total number of steps at the beginning, followed by step-by-step verbal guidance for positioning arms, legs, torso, and face. The instructions were pre-recorded in a clear, neutral tone to ensure uniform delivery across participants. The detailed audio instruction scripts are provided in Appendix~\ref{appendix-audio}.

\begin{figure*}
  \centering
  \includegraphics[width=\textwidth]{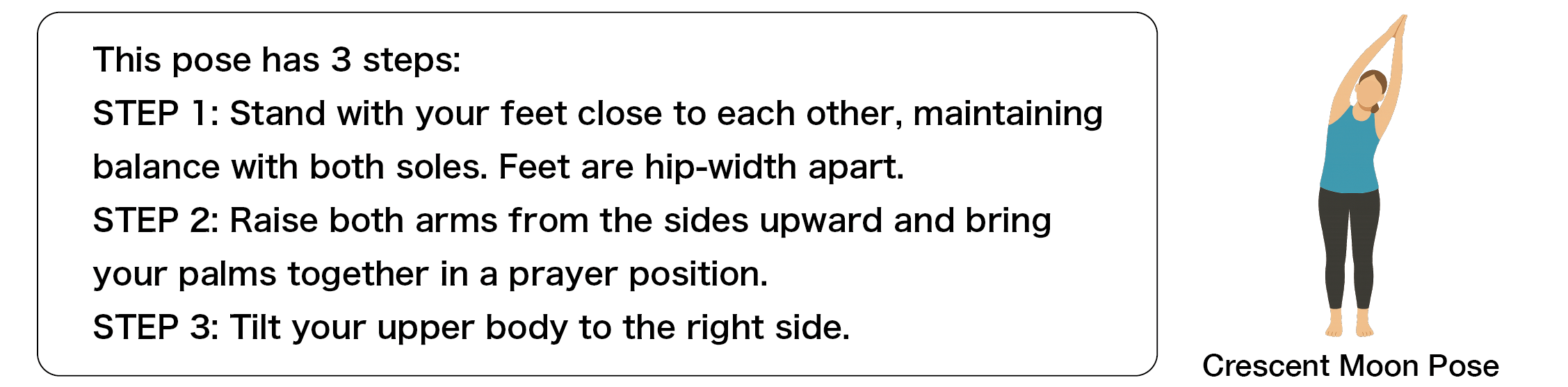}
  \caption{Example of audio instruction format: First announcing the total number of steps, then explaining each step individually for the Crescent Moon Pose. This pose has three steps: STEP 1: Stand with your feet close to each other, maintaining balance with both soles. Feet are hip-width apart. STEP 2: Raise both arms from the sides upward, and bring your palms together in a prayer position. STEP 3: Tilt your upper body to the right side.}
  \Description{Figure 6 illustrates the format of audio-based step-by-step instructions using the Crescent Moon Pose as an example. On the left, a text box states that the pose consists of three steps, each written in bold: Step 1 instructs the user to stand with feet hip-width apart and balanced; Step 2 involves raising both arms and joining the palms; Step 3 requires tilting the upper body to the right. On the right side of the figure, there is an illustration of a woman performing the Crescent Moon Pose, with arms raised and body leaning to one side.}
  \label{fig:audio}
\end{figure*}

\begin{figure*}
  \centering
  \includegraphics[width=\textwidth]{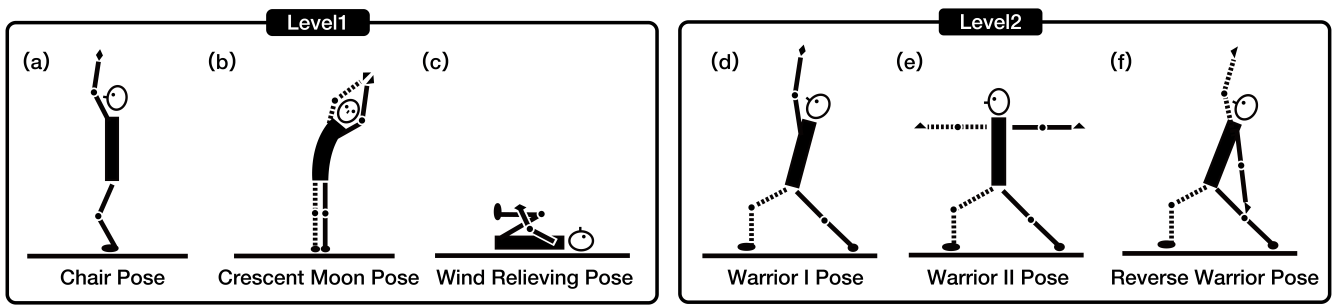}
  \caption{2D tactile graphics representing Yoga poses at two difficulty levels: Level 1 includes (a) Chair Pose, (b) Crescent Moon Pose, and (c) Wind Relieving Pose; Level 2 includes (d) Warrior I Pose, (e) Warrior II Pose, and (f) Reverse Warrior Pose}
  \Description{Figure 7 presents 2D tactile graphics depicting six yoga poses across two difficulty levels. Level 1 on the left includes (a) Chair Pose, (b) Crescent Moon Pose, and (c) Wind Relieving Pose. Level 2 on the right includes (d) Warrior I Pose, (e) Warrior II Pose, and (f) Reverse Warrior Pose. Each pose is illustrated using simplified stick-figure representations with solid black lines for limbs, circles for joints and heads, and no directional arrows.}
  \label{fig:2d_yoga}
\end{figure*}

\subsubsection{Design of 2D Tactile Graphics}
The 2D tactile graphics were created based on the design methods for human body representation design methods in Holloway et al \cite{Holloway2022}. We used the images depicted in Figure \ref{fig:2d_yoga} to distinguish between the left and right limbs. Additionally, a horizontal line representing the floor was placed at the bottom to serve as a reference point for position and posture.

For the presentation method of the 2D tactile graphics used in the experiment, we conducted a collaborative evaluation with visually impaired participant P0 to compare two types of tactile display media: Swell Paper and a braille display (Dot PAD / DPA320ADot). When evaluating the tactile perceptibility using graphics created on both media, we confirmed that the Swell Paper was superior in representing contour details. P0 preferred Swell Paper, since Dot PAD placed constraints on resolution (low resolution) and interaction (inability to touch two graphics simultaneously). \ken{Based on these comparative results, we adopted 2D tactile graphics printed on A4-sized Swell Paper, with each figure having an approximate size of 15~cm~$\times$~15~cm.}

\subsection{Participants}
\begin{table}[htbp]
\centering
\caption{Demographic information of the participants.}
\Description{Table 1 shows demographic information of 10 participants, including gender, age, years since becoming blind, and frequency of exercise. Ages range from 21 to 57, and exercise habits vary from “Not at all” to “4–5 times a week.”}
    \label{tab:participant_info}
\begin{tabular}{ccccc}
\toprule
ID & Gender & Age & Blind Since & Exercise Frequency \\
\midrule
P1 & Male & 55 & 8 & Once a month \\
P2 & Male & 21 & 3 & 4-5 times a week \\
P3 & Female & 56 & 0 & Not at all \\
P4 & Female & 44 & 9 & Once a month \\
P5 & Female & 50 & 3 & Not at all \\
P6 & Male & 33 & 0 & Not at all \\
P7 & Female & 57 & 53 & Once a week \\
P8 & Male & 54 & 12 & 3-4 times a week \\
P9 & Male & 45 & 18 & 3-4 times a week \\
P10 & Male & 50 & 30 & Once a month \\
\bottomrule
\end{tabular}
\end{table}

As shown in Table \ref{tab:participant_info}, we recruited 10 blind participants (7 males, 3 females) aged between 21 and 56 years (M = 46.5, SD = 10.9). Recruitment was conducted through an electronic newsletter for visually impaired individuals, and participants received a compensation of approximately 50 USD along with travel expenses. \ken{During a preliminary interview, five participants (P1, P3, P6, P8, P9) reported having no prior experience with yoga, while the remaining five (P2, P4, P5, P7, P10) had tried it only a few times in the past. Regarding calisthenics, all participants mentioned having only brief experience during physical education classes in their school years. In addition, all participants had some prior experience with tactile graphics and 3D models through previous educational contexts. Both Study~1 and Study~2 were conducted on the same day for each participant.} The study received approval from the Ethics Review Committee.

\subsection{Task and Procedure}
Prior to the main study, participants received a total of 5-minute introduction to the methods using sample materials. For the audio method, participants listened to instructions describing a standing pose with both arms raised. 
For the 2D tactile graphic, participants explored a tactile representation of a standing pose under researcher guidance, identifying key elements such as arm and leg positions, face direction, palm orientation, and joint positions. 
For the 3D model, participants handled a model of the same standing pose, confirming the direction of the face, limb positions, and palm orientations.

Participants were presented with six yoga poses (three from Level 1 and three from Level 2), each described using one of the three methods respectively in randomized order: audio instructions, 2D tactile graphics, and 3D models (Table \ref{tab:user_study_1}). \ken{Each participant experienced all three methods across different poses, but only one method was provided for each pose.} Participants were allowed to ask questions during tactile exploration, and the number of inquiries was recorded. Participants were also free to alternate between touching the material and physically trying the pose to aid understanding.

After understanding each pose, participants were asked to reproduce it. Following reproduction, feedback was provided regarding the correctness of the pose. The time required from initial understanding to reproduction was recorded. To enable objective evaluation, photographs of participants’ reproduced poses were collected.  These reproductions were evaluated using a scoring rubric with a maximum of 8 points: head (1 point), pelvis (1 point), left arm (1 point), right arm (1 point), left leg (1 point), right leg (1 point), left wrist (0.5 point), right wrist (0.5 point), left ankle (0.5 point), and right ankle (0.5 point). Although this was not a standardized measure, the weighting scheme was inspired by the joint-specific criteria used by Dias et al.~\cite{Dias2019} to evaluate movement reproduction. This ensured a consistent and transparent method for assessing pose accuracy across participants.

This session lasted approximately 30-45 minutes.
Upon completing all poses, a semi-structured interview was conducted. As shown in Figure \ref{fig:us1_result}, the interview consists of questions ranging from Q1 to Q6 (A: Audio, B: 2D, C: 3D), with free comments. This session lasted approximately 15 - 30 minutes, with the entire study taking around 1 hours.

\subsection{Results}

\begin{table}[htbp]
\centering
\caption{Results for Average Time, Scores, and Number of Questions}
\Description{Table 2 compares average completion time (in seconds), accuracy score (out of 8), and number of clarification questions asked across three instructional methods—Audio, 2D, and 3D—for both beginner (Level 1) and intermediate (Level 2) yoga poses.}
\setlength{\tabcolsep}{2pt}
\begin{tabular}{l|ccc|ccc|ccc}
\hline
& \multicolumn{3}{c|}{\makecell{\textbf{Average Time} \\ \textbf{(seconds)}}}
 & \multicolumn{3}{c|}{\textbf{Score (max: 8)}} 
 & \multicolumn{3}{c}{\makecell{\textbf{Number of} \\ \textbf{Questions\textsuperscript{*}}}}\\
\cline{2-10}
 & Audio & 2D & 3D & Audio & 2D & 3D & Audio & 2D & 3D \\
\hline
Yoga Level1 & 68.6 & 155 & 48.2 & 7.0 & 6.95 & 7.9 & 1.8 & 3.7 & 0.5 \\
Yoga Level2 & 98.6 & 135.2 & 72 & 6.3 & 6.55 & 7.5 & 2.1 & 2.1 & 0.9 \\
\hline
\end{tabular}
\caption*{\small \textsuperscript{*}For Audio, the number indicates the times the audio was replayed.}
\end{table}

\subsubsection{\textbf{Analysis of Objective Indicators}}

In this section, we report the results of User Study 1, focusing on participants’ performance in reproducing yoga poses across three instructional methods. We present our analysis from three perspectives: completion time, pose accuracy, and clarification behaviors including questions and audio replays. In addition, we examine participants’ subjective ratings using a 7-point Likert scale and conduct a qualitative analysis to highlight both positive feedback and suggestions for improvement. Together, these measures provide a comprehensive view of how 3D models, 2D tactile graphics, and audio instructions supported blind participants’ understanding of yoga poses.

\textbf{Task Completion Time:}  
For Level 1 poses, Wilcoxon signed-rank tests with Bonferroni correction ($\alpha = 0.0167$) revealed that 3D models were significantly faster than 2D tactile graphics ($p = 0.002$). No significant differences were found between 3D models and audio instructions ($p = 0.1602$) or between audio instructions and 2D tactile graphics ($p = 0.0195$). For Level 2 poses, no significant differences were observed among the methods ($p = 0.0672$). Nevertheless, the average completion times consistently followed the order of 3D models being the fastest, followed by audio instructions, and then 2D tactile graphics.

\textbf{Reproduction Accuracy Scores:}  
Accuracy was assessed using the 8-point rubric described in Section~4.3. For Level 1 poses, no significant differences were observed among the methods ($p = 0.289$). For Level 2 poses, Friedman tests indicated a significant effect of condition ($p = 0.026$). Post-hoc pairwise comparisons with Bonferroni correction ($\alpha = 0.0167$) showed that 3D models outperformed 2D tactile graphics ($p = 0.014$), while no significant differences were found between 3D models and audio instructions or between audio instructions and 2D tactile graphics. Overall, participants achieved the highest reproduction accuracy with 3D models, which outperformed both audio instructions and 2D tactile graphics across pose levels.

\textbf{Clarification Questions and Audio Replays:}  
Participants using 3D models asked very few clarification questions, averaging 0.5 for Level 1 and 0.9 for Level 2. Notably, 80\% of participants were able to understand the poses without any clarification, indicating strong clarity despite the increased complexity of Level 2.
In contrast, participants using 2D tactile graphics asked an average of 3.7 questions for Level 1 and 2.1 for Level 2. Most participants reported that imagining three-dimensional shapes from flat diagrams was difficult, and that it took time to form a mental image.
For audio instructions, participants frequently replayed the recordings, averaging 1.9 replays for Level 1 and 2.1 for Level 2. Most participants noted that multiple replays were necessary to fully understand the instructions. P1 specifically pointed out the difficulty of navigating the audio, stating, \textit{``I couldn't replay just the part I was unsure about, so I had to listen to the entire audio again and again. It made it hard to go at my own pace.''}

\subsubsection{\textbf{Usability and User Experience Evaluation}}
\begin{figure*}
  \centering
  \includegraphics[width=\textwidth]{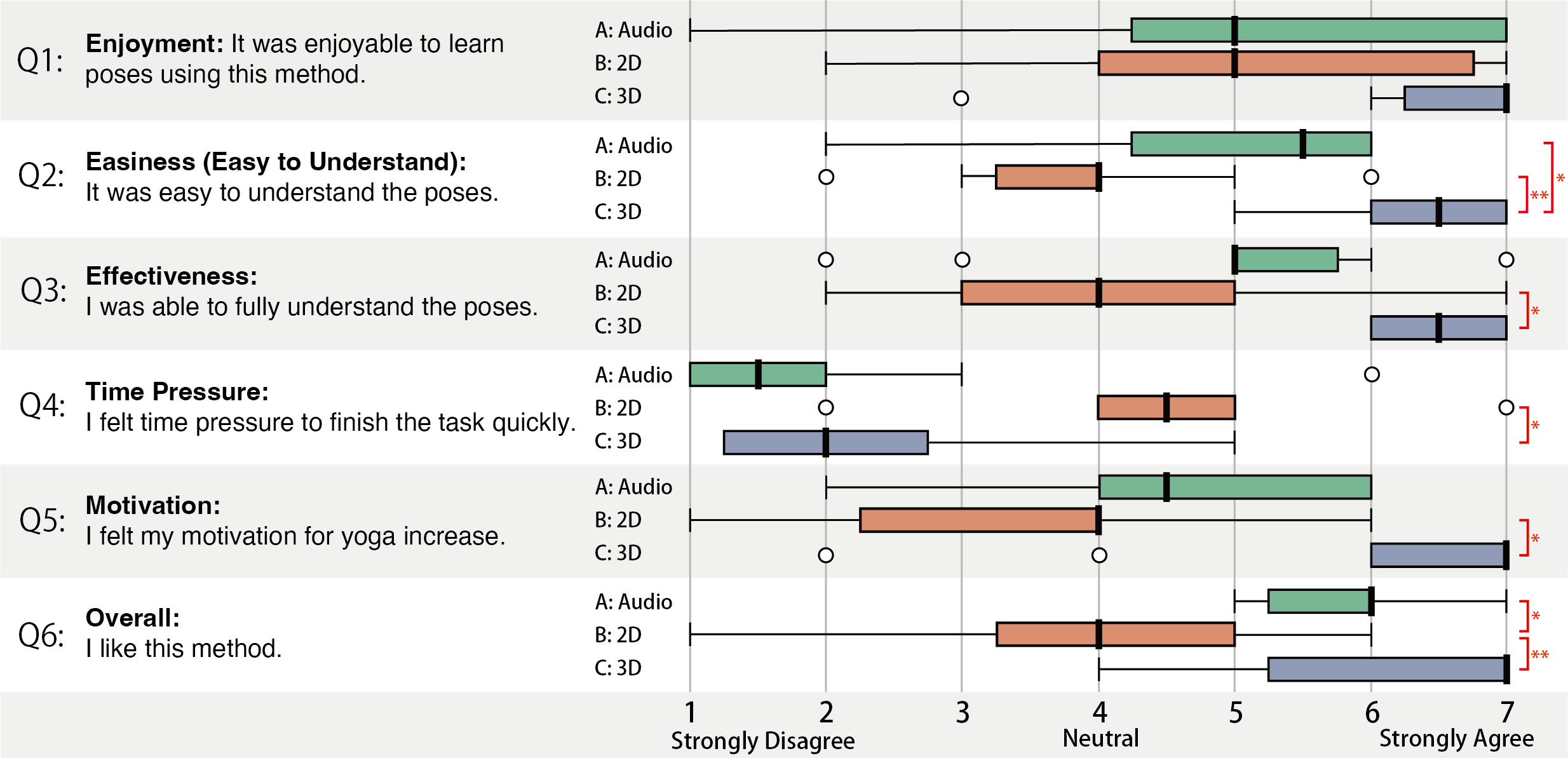}
  \caption{Questionnaire results comparing user evaluations of yoga-pose learning methods among three different approaches: audio instructions (A: Audio), 2D tactile graphics (B: 2D), and 3D models (C: 3D). Assessments included six aspects: Q1 (Enjoyment: it was enjoyable to learn poses using this method), Q2 (Ease of Understanding: It was easy to understand the poses), Q3 (Effectiveness: I was able to fully understand the poses), Q4 (Time Pressure: I felt time pressure to finish the task quickly), Q5 (Motivation: I felt my motivation for yoga increase), and Q6 (Overall: I like this method). All responses were recorded on a 7-point Likert scale ranging from 1 (Strongly Disagree) to 7 (Strongly Agree). Asterisks indicate statistically significant differences based on pairwise Wilcoxon signed-rank tests with Bonferroni correction: * $p < .0167$, ** $p < .0033$.}
  \Description{Figure 8 presents box plots comparing user ratings across three yoga-pose learning methods—audio instructions (A), 2D tactile graphics (B), and 3D models (C)—across six evaluation aspects: Q1 (Enjoyment), Q2 (Ease of Understanding), Q3 (Effectiveness), Q4 (Time Pressure), Q5 (Motivation), and Q6 (Overall). Ratings were collected using a 7-point Likert scale from 1 (Strongly Disagree) to 7 (Strongly Agree).}
  \label{fig:us1_result}
\end{figure*}

In Figure~\ref{fig:us1_result}, we report the results of the six seven-point Likert items (Easiness, Effectiveness, Time Pressure, Motivation, Overall, and Enjoyment). For Likert items, a median score of five or higher indicates that participants generally responded positively. Friedman tests followed by Wilcoxon signed-rank tests with Bonferroni correction ($\alpha = 0.0167$) showed significant differences between methods across five evaluation criteria—Easiness, Effectiveness, Time Pressure, Motivation, and Overall—but not for Enjoyment. 
Specifically, 3D models were rated significantly higher than 2D tactile graphics across all five of these criteria (all $p < 0.0167$). For Easiness, 3D models also outperformed audio instructions ($p = 0.0097$). For Overall, audio instructions received significantly higher ratings than 2D tactile graphics ($p = 0.0098$). 

These results suggest that 3D models generally provided the clearest and most intuitive support for understanding yoga poses in the context of this study, particularly when compared with 2D tactile graphics and audio instructions. They also outperformed audio instructions in easiness, indicating their potential to reduce perceived memory demands by allowing participants to directly and simultaneously perceive spatial relationships rather than reconstructing them from sequential verbal instructions. Although no significant differences were observed for enjoyment, audio instructions still received higher ratings than 2D tactile graphics in overall evaluation, underscoring the continuing value of structured verbal guidance.

\subsubsection{\textbf{Qualitative Analysis}}  
\textbf{Positive Feedback:}  
All participants reported that the 3D tactile models helped them clearly and intuitively understand the overall shape of each yoga pose. 
Unlike 2D graphics or verbal descriptions, which often required mentally piecing together disconnected body parts, the 3D models provided a complete, coherent posture through direct touch. 
This made it easier to form a mental image of the pose without relying on guesswork. 
For example, P1 remarked, \textit{``The 3D model was accurate in shape, and I could immediately imagine the pose when I touched it. Compared with other methods, it required less guessing and was easier to understand.''} 
This suggests that whole-body tactile representations reduce the cognitive effort needed to reconstruct poses from partial or symbolic cues. 
Similarly, P3 stated that the pose \textit{``was conveyed as it was,''} indicating that the model closely matched the mental image they formed through touch.

In addition to the overall posture, six participants (P2--P4, P7, P9, P10) noted that the 3D models helped them identify fine details---such as hand orientation, face direction, and ankle angles---that were often difficult to interpret using 2D diagrams or verbal instructions. 
P3 explained that small features like the nose and palms clarified the figure’s orientation: \textit{``I could tell whether the palm was facing up or down... I could also feel the nose and ears, which made the overall image much clearer.''} 
P7 described a similar learning process, starting with the overall shape and then moving to specific parts: \textit{``By first grasping the overall form and then checking the finer parts like arms and hands, I could understand the pose more accurately.''} 
These comments highlight how the 3D models supported an exploratory approach where users first established a general understanding of the pose before examining individual features in detail.

Many participants also described feeling more confident and motivated when using the 3D models. 
Six participants (P2--P5, P8, P10) said that the tactile clarity of the models made the activity feel more accessible and less intimidating. 
For instance, P4 explained that being able to confirm the shape through touch gave them a sense of capability: \textit{``It made me feel like I could do it.''} 
Similarly, P8 shared that the model made the task feel easier and more approachable: \textit{``I felt this level could be understood without effort, so I felt safe to try it.''} 
These comments suggest that physical models not only enhance understanding but also lower psychological barriers to participation---particularly important in learning activities involving movement.

Six participants (P2--P5, P8, P10) noted that the clarity of the 3D models made the activity feel more approachable. 
Being able to confirm the pose through touch gave them a greater sense of confidence. 
P4 remarked, \textit{``It made me feel like I could do it,''} suggesting that tactile confirmation helped turn uncertainty into a sense of ability. 
P8 said, \textit{``I felt this level could be understood without effort, so I felt safe to try it,''} indicating that clear guidance reduced hesitation. 
These responses suggest that the 3D models not only improved understanding but also made participants feel more willing and comfortable to try the movements.

Finally, three participants (P6, P7, P10) emphasized the value of being able to freely rotate the models during exploration. 
This made it easier to align the model with their own body and imagine how to reproduce the pose. 
P6 said, \textit{``With 2D tactile graphics, I had to mentally flip left and right, which made it difficult to understand the posture. But with the 3D model, I could freely change its orientation to match my own.''} 
This highlights how physical manipulability can reduce mental effort and help users relate the pose to their own body. 
P7 also pointed out the benefit of the model’s small size and ease of handling: \textit{``It is difficult to freely touch a yoga instructor’s body, but the 3D model was small in size, fit well in my hands, was easy to handle, and allowed me to check it repeatedly.''} 
The model’s portability and tactile clarity allowed participants to explore at their own pace and revisit unclear parts as needed.

\textbf{Negative Feedback and Suggestions:}  
Despite the overall positive reception, some participants found it challenging to understand how to move into the final pose when only the completed posture was presented. Four participants (P1, P4, P5, P10) noted that the lack of information about transitional steps required them to mentally reconstruct the movements, which led to uncertainty and increased effort. P1 explained, \textit{``The final pose was clear, but the movements leading up to it were not conveyed. As a result, I had to imagine and adjust on my own.''} P4 similarly remarked, \textit{``I could understand the final pose, but since the transition was not clear, I felt uncertain whether I was moving correctly from the start.''} This feedback suggests that some users may benefit from additional guidance to better understand how to reach the final posture.

To supplement the 3D models, four participants (P5--P7, P9) suggested combining them with audio instructions. They felt that a multimodal approach could provide sequential guidance and reduce the burden of remembering or inferring transitions. P5 commented, \textit{``While audio does not convey small details, it is good that I can follow the explanation while moving. If I could learn using both the 3D model and audio together, it would be very helpful.''} P7 proposed a structured approach: \textit{``It would be good to first understand the overall form with the 3D model and then listen to the step-by-step explanations with audio.''} This perspective highlights the potential of integrating tactile and auditory information to support both spatial understanding and temporal sequencing.

Two participants (P1, P5) also expressed a desire for corrective feedback during practice. Without confirmation of whether their pose was accurate, they felt uncertain about their performance. P1 said, \textit{``When practicing in a group, it would be reassuring if I could confirm whether my pose was correct.''} This suggests a need for reassurance through external feedback. P5 stated, \textit{``It is difficult for me to notice small differences by myself, so I would like to be told whether my pose was correct.''} This points to the difficulty of self-assessing subtle errors without support. These comments suggest that self-guided learning may benefit from mechanisms that allow users to verify the accuracy of their movements.

%% file: SECTIONS/5_Study2.tex
\section{USER STUDY 2: Comparing 3D Models With 2D Tactile Graphics for Comprehending Calisthenic Movements}

This study was conducted with the same 10 blind participants introduced in Section 4.2. Since the purpose of User Study 2 was to compare the effectiveness of different instructional methods with the same participant group, no additional Participants section is included here; all demographic information is provided in Section 4.2.

\begin{figure*}
  \centering
  \includegraphics[width=\textwidth]{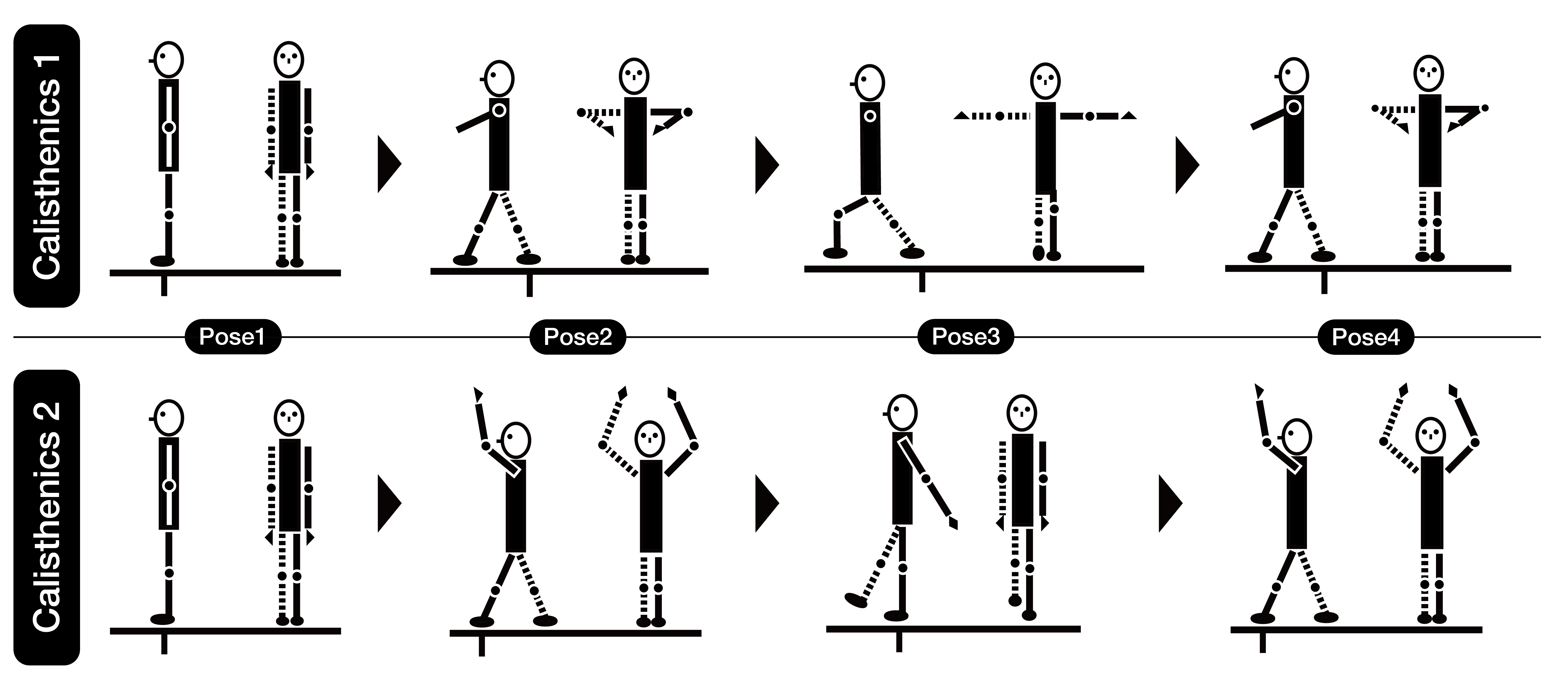}
  \caption{2D tactile graphics (Calisthenics 1 and 2): Each pose is presented from two different perspectives (frontal view and lateral view).}
  \Description{Figure 9 shows 2D tactile graphics illustrating two calisthenic movement sequences, Calisthenics 1 and Calisthenics 2, each consisting of four poses labeled Pose 1 to Pose 4. Each pose is represented by two stick-figure illustrations—one from a lateral view and one from a frontal view—positioned side by side. All figures are designed using a tactile-friendly format: the arms and legs are visually distinguished by using solid lines for arms and dotted lines for legs. Circular shapes are used for joints and heads. The poses are arranged horizontally from left to right, with triangular arrows indicating progression through the movement sequence. Calisthenics 1 is shown in the top row, and Calisthenics 2 is shown in the bottom row.}
  \label{fig:yoga_2d}
\end{figure*}

\subsection{Apparatus for Comparison}

The 2D tactile graphics were created with reference to designs similar to those used in User Study 1 as shown in Figure \ref{fig:yoga_2d}. To clearly differentiate between left and right, the left arm and left leg were represented with dotted lines, while the right arm and right leg were depicted with solid lines. A horizontal line was drawn to indicate the floor position, and additionally, a vertical line extending downward from the floor was added as a reference mark for the initial standing position. This marker made it possible to express which direction, forward or backward, the feet would move. Furthermore, to accurately convey the forward and backward movements of the feet and the depth of the arms in calisthenic poses, each pose was represented from two perspectives: frontal and lateral views. Detailed methods for creating the 3D models are described in Section 3.

\ken{User Study~1 used audio instructions for yoga poses, which have standardized forms and reference scripts reviewed by instructors. In contrast, the calisthenic exercises selected for this study presented different challenges. Unlike yoga, which can be described through step-by-step verbal guidance, these exercises are typically demonstrated visually to sighted learners, making it difficult to create equivalent audio scripts for their continuous and connected movements. Moreover, verbal instructions often fail to convey movement style, subtle nuances, and motion fluidity~\cite{Holloway2022}. These limitations were expected to be more pronounced in four-step calisthenic sequences; therefore, audio methods were not used in this study.}

\subsection{Task and Procedure}
User Study 2 aimed to compare the effectiveness of 2D tactile graphics and 3D models for conveying dynamic calisthenic poses. The experiment randomly presented two types of calisthenic sequences (Table \ref{tab:user_study_2}), each consisting of four consecutive poses, using either 2D tactile graphics or 3D models. The second and fourth poses were identical, although this information was not disclosed to participants beforehand.

Prior to the main study, participants received 5-10 minutes of instruction on each method using sample materials. For 2D tactile graphics, participants explored graphics of standing poses under researcher guidance, identifying critical elements such as arm and leg positions, facial orientation, palm direction, and joint locations. They were also informed that each pose was represented from two perspectives (frontal and lateral) and were instructed about the reference line indicating the floor. For 3D models, participants handled models of standing poses, examining facial orientation, limb positions, and palm directions through touch. Additionally, the significance of reference markers positioned in front of the toes was explained.

In the main study, after comprehending the four sequential movements, participants were asked to physically reproduce the movement sequence. As in User Study 1, their performance was evaluated using an 8-point scoring rubric, which assessed the accuracy and completeness of the reproduced movements. The time required from understanding to reproducing the movement was also measured, and photographs of their poses were collected for objective evaluation. This session lasted approximately 20 minutes.
Upon completing all poses, a semi-structured interview was conducted. As shown in Figure \ref{fig:us2_result}, the interview consisted of questions ranging from Q1 to Q6 (A: 2D, B: 3D) and free comments. This session lasted approximately 20 minutes, with the entire study taking around 40 minutes.

\subsection{Results}

\begin{table}[htbp]
\centering
\caption{Results for Calisthenic Movements (2D vs 3D)}
\Description{Table 3 summarizes performance metrics for Calisthenics 1 and 2 across 2D and 3D conditions. 3D models yielded faster times, higher scores, and fewer clarification questions compared to 2D graphics.}
\setlength{\tabcolsep}{3pt}
\begin{tabular}{l|cc|cc|cc}
\hline
 & \multicolumn{2}{c|}{\textbf{Time (seconds)}} 
 & \multicolumn{2}{c|}{\textbf{Score (max:8)}} 
  & \multicolumn{2}{c}{\makecell{\textbf{Number of} \\ \textbf{Questions}}}\\

\cline{2-7}
 & 2D & 3D & 2D & 3D & 2D & 3D \\
\hline
Calisthenic 1 & 474.6 & 218.0 & 7.4 & 7.6 & 4.2 & 1.0 \\
Calisthenic 2 & 378.8 & 313.6 & 7.3 & 7.7 & 4.6 & 2.2 \\
Overall       & 426.7 & 265.8 & 7.35 & 7.65 & 4.4 & 1.6 \\
\hline
\end{tabular}
\end{table}

\subsubsection{\textbf{Analysis of Objective Indicators}}

In this section, we report the results of User Study 2, focusing on participants’ performance in reproducing sequential calisthenic movements across two instructional methods. We present our analysis from three perspectives: completion time, pose accuracy, and clarification behaviors including questions. In addition, we examine participants’ subjective ratings using a 7-point Likert scale and conduct a qualitative analysis to capture positive feedback and suggestions for improvement. Together, these measures provide a comprehensive view of how 3D models and 2D tactile graphics supported blind participants’ understanding of calisthenic sequences.

\textbf{Task Completion Time:} 
Wilcoxon signed-rank test results demonstrated that the 3D models allowed significantly faster movement recognition and reproduction compared to the 2D tactile graphics ($p = 0.0059$). Nine of the ten participants (all except P5) achieved time reduction with the 3D models.

\textbf{Reproduction Accuracy Scores:} 
Although no significant difference was found between 3D models and 2D tactile graphics ($p = .1677$), both methods resulted in average scores exceeding 7 points, suggesting that participants were generally able to understand and reproduce the calisthenic movements with few major errors. With sufficient time, both instructional formats allowed participants to accurately grasp the sequential motions. Most of the observed errors were minor, primarily involving incorrect arm positions.

\textbf{Number of Questions:} 
For calisthenic movements, the participants asked fewer questions when using 3D models compared to 2D tactile graphics. Participants using 3D models asked an average of 1.6 questions, while those using 2D tactile graphics asked an average of 4.4 questions. 
Furthermore, 60\% of participants required no clarification questions at all when using 3D models, demonstrating the inherent clarity of this representation method. In contrast, all participants using 2D tactile graphics needed to ask at least one question, with some participants (P5 and P6) requiring substantial clarification (8 and 15 questions respectively).
The difference in question count suggests that 3D models provide a more intuitive and comprehensive understanding of calisthenic movements compared to 2D tactile graphics.

\subsubsection{\textbf{Usability and User Experience Evaluation}}
\begin{figure*}
  \centering
  \includegraphics[width=\textwidth]{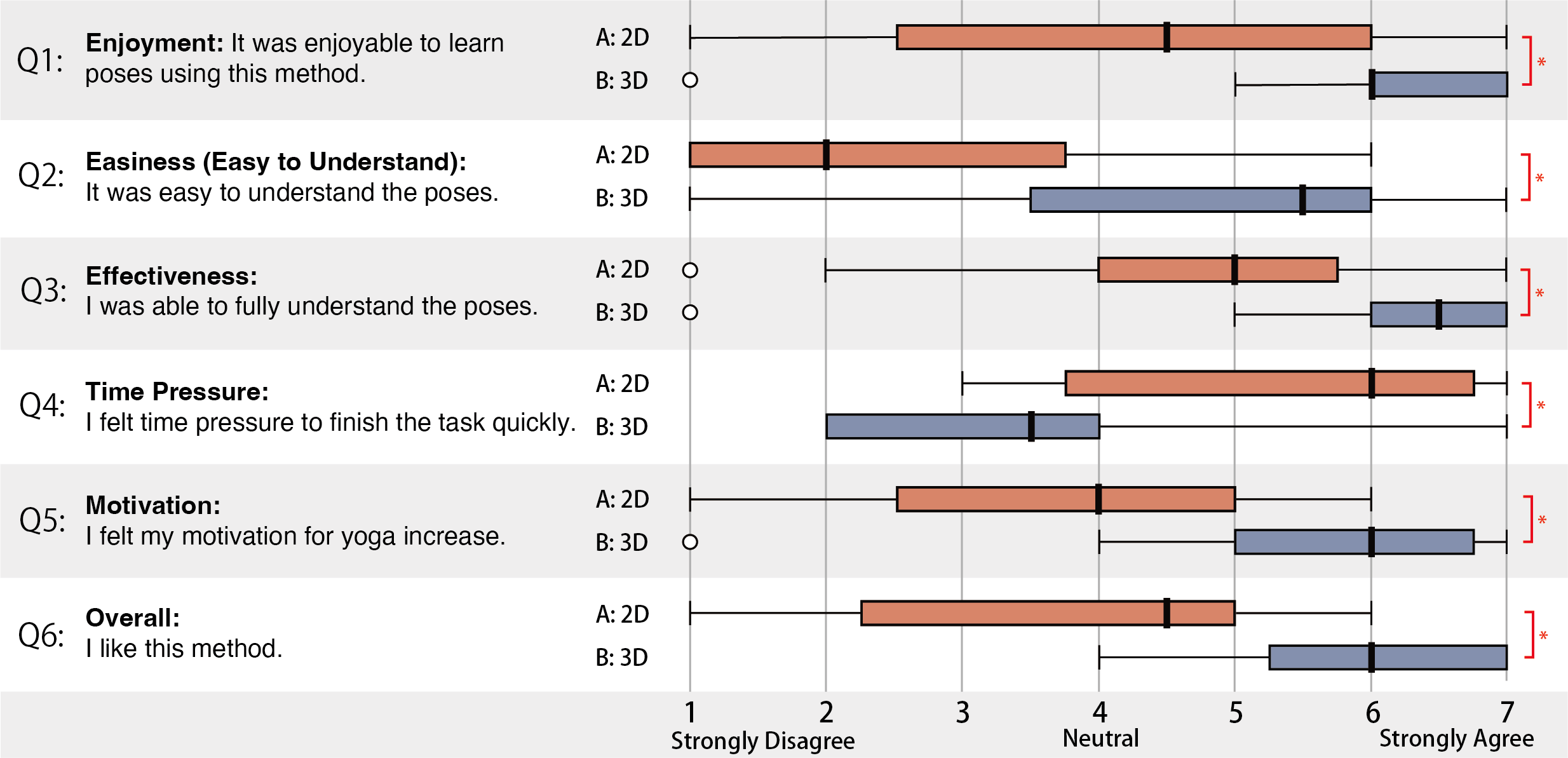}
  \caption{Questionnaire results comparing user evaluation of learning methods for calisthenic poses learning methods between 2D tactile graphics (A: 2D) and 3D models (B: 3D). The assessment included six aspects: Q1 (Enjoyment: it was enjoyable to learn poses using this method), Q2 (Ease of Understanding: It was easy to understand the poses), Q3 (Effectiveness: I was able to fully understand the poses), Q4 (Time Pressure: I felt time pressure to finish the task quickly), Q5 (Motivation: I felt my motivation for calisthenics increase), and Q6 (Overall: I like this method). All responses were recorded on a 7-point Likert scale ranging from 1 (Strongly Disagree) to 7 (Strongly Agree). Asterisks indicate statistically significant differences based on Wilcoxon signed-rank tests: * $p < .05$, ** $p < .01$.}
  \Description{Figure 10 presents box plots comparing user evaluations of two learning methods for calisthenics poses: 2D tactile graphics (A) and 3D models (B), across six aspects: Q1 (Enjoyment), Q2 (Ease of Understanding), Q3 (Effectiveness), Q4 (Time Pressure), Q5 (Motivation), and Q6 (Overall). Ratings are plotted on a 7-point Likert scale from 1 (Strongly Disagree) to 7 (Strongly Agree).}
  \label{fig:us2_result}
\end{figure*}

In Figure~\ref{fig:us2_result}, we report the results of the six seven-point Likert items (Easiness, Effectiveness, Time Pressure, Motivation, Overall, and Enjoyment). A median score of five or higher indicates that participants generally responded positively. Wilcoxon signed-rank tests were conducted to compare participants' subjective ratings between 2D tactile graphics and 3D models across all six items. The analysis revealed statistically significant differences in all six questionnaire items (Q1–Q6; all p < 0.05)

Specifically, 3D models were rated higher than 2D tactile graphics in Enjoyment ($p = 0.041$), Easiness ($p = 0.014$), Effectiveness ($p = 0.028$), Time Pressure ($p = 0.017$), Motivation ($p = 0.034$) and Overall ($p = 0.010$).For Easiness, participants emphasized that the ability to freely adjust the orientation of the 3D models made it much easier to understand the posture. For Effectiveness, many participants highlighted that they could immediately recognize the form and reproduce the pose. With regard to Time Pressure, most participants reported that 3D models reduced perceived cognitive load compared to interpreting 2D diagrams. Motivation scores also reflected that participants felt more willing to attempt the movements when using 3D models. 

Taken together, these results demonstrate that 3D models consistently provided clearer and more intuitive support for understanding calisthenic sequences than 2D tactile graphics. Importantly, the consistent and statistically significant advantages observed across all evaluation criteria indicate that 3D models are effective not only for static postures, such as yoga poses examined in Study 1, but also for dynamic movement sequences, such as the calisthenic tasks in Study 2. This highlights their strong potential as an effective and scalable approach for supporting blind individuals in learning both static and dynamic physical activities.

\subsubsection{\textbf{Qualitative Analysis}}

\textbf{Positive Feedback:}  
Extending from static pose comprehension in User Study~1, most participants in User Study~2 reported that 3D models were also helpful for learning sequences of movements. They noted that the models helped them understand the overall flow and see how each part of the body should be positioned.

Five participants (P1, P4, P7, P8, P10) said they could quickly and clearly understand each pose in the sequence. P1 stated, \textit{``With 2D, I had to combine the front and side views in my head, which was difficult, but with the 3D model, I could instantly understand the shape of the pose.''} This shows that 3D models made it easier to understand poses without needing to combine different views mentally. P4 similarly noted, \textit{``With the 3D model, I could quickly understand and reproduce the posture,''} suggesting that being able to touch the whole shape helped them remember the pose. P10 added, \textit{``By rotating the model to align with my own direction, I could imagine the flow of movements,''} showing that being able to change the direction of the model helped them picture how to move.

Four participants (P4, P7, P9, P10) also found the models useful for understanding how the poses connected. They explained that the models helped them picture how the hands and feet should move from one pose to the next. P7 said, \textit{``I could first understand the overall sequence of movements, and then confirm the detailed parts,''} suggesting they used the model step by step. P10 remarked, \textit{``I was able to imagine how the hands and feet moved between each posture,''} showing that the model helped them connect each pose in their mind.

All participants also mentioned that the tactile reference points on the models were useful. P1 said, \textit{``The reference points were very helpful; I could tell which leg was in front,''} meaning that small markers made it easier to understand the shape. P7 said, \textit{``Starting with the reference points helped me understand the direction of movement,''} and P9 added, \textit{``Markers that are hard to distinguish in 2D were clear in 3D. For diagonal movements, 3D would be necessary to understand them,''} showing how 3D shape helped them feel direction and motion more clearly.

\textbf{Negative Feedback and Suggestions:} 
While most participants were able to follow the pose sequence, four participants (P2, P5, P6, P8) noted that keeping track of the movement order and recalling it during execution was mentally demanding. P5 said, \textit{``I understood the sequence of four poses, but remembering the order was difficult, and I worried about reproducing it correctly.''} P8 said, \textit{``If the number of poses increases, it will become harder to remember,''} raising concerns about the increased memory load for longer sequences.

To help with memory and make transitions easier to follow, five participants (P1, P5, P6, P9, P10) suggested that audio explanations, provided step by step, could improve learning when used alongside the 3D models. P6 said, \textit{``If there had been audio to explain how to move between poses, it would have been easier to follow,''} noting that verbal instructions could guide transitions. P5 remarked, \textit{``Since I'm used to audio explanations, combining them with touch would make it easier to remember,''} indicating that using both tactile and auditory information together could reinforce recall and improve learning.

Two participants (P1, P10) also expressed a desire for feedback to confirm whether their body positions matched the intended poses. P1 said, \textit{``It would be reassuring if I could confirm whether my pose is correct,''} highlighting the desire for confirmation during self-practice. P10 noted, \textit{``The model helps me understand the shape, but it's hard to know if I'm doing it right. If there were explanations while moving, I could confirm whether my pose is correct,''} emphasizing the benefit of receiving real-time feedback while performing the movements.

In addition, two participants (P4, P7) suggested adding cues to make the models easier to identify and orient. P4 said, \textit{``When the 3D models were placed on the table, there were times when I was unsure which pose I was touching, so adding numbers would help,''} pointing to confusion when identifying poses by touch. P7 remarked, \textit{``As I rotated the model, there were moments when I lost track of which way was the front, so a fixed tactile marker on the model to indicate its front side would be necessary when the movement involves rotation,''} suggesting that consistent orientation markers could improve usability.

%% file: SECTIONS/7_Discussion.tex
\section{DISCUSSION}

\subsection{Understanding Poses and Movements through 3D Tactile Models}

Existing non-visual instructional methods, such as verbal descriptions and 2D tactile graphics, often struggle to convey whole-body spatial relationships and fine-grained limb orientations. Our findings show that 3D tactile models address these limitations by enabling blind users to comprehend full-body poses through direct physical exploration. Participants could perceive both overall structure and fine-grained details—such as joint angles, hand orientation, and body direction—more clearly than with audio descriptions or 2D tactile graphics. By reducing the need for mental reconstruction, the models enabled participants to form an accurate mental image of each posture, and many reported greater confidence in reproducing the poses after tactile exploration.

This approach also expands the range of postures that can be effectively taught through tactile means. Unlike prior systems that focused on standing poses or simplified body shapes, our models supported the exploration of more complex configurations, including the Wind Relieving Pose performed while lying on the floor. Such non-standing postures are difficult to capture through skeletal tracking systems like Kinect~\cite{Rector2017, RectorYoga2014, RectorYoga2013}, yet were effectively conveyed in tactile form in our study.

In addition, prior work such as Holloway et al.~\cite{Holloway2022} highlighted challenges in conveying spatial relationships between limbs using 2D tactile graphics, due to limitations in resolution and tactile area. Our 3D models addressed this issue by enabling users to explore full-body structures through touch, facilitating understanding of limb relationships in poses where arms and legs are bent, overlapped, or close to the torso.

The models were also effective in sequential learning contexts. In User Study~2, participants used a series of four models to learn calisthenic-style movements. Despite their static nature, many were able to reconstruct the intended motion by comparing differences between adjacent poses. These findings suggest that a well-designed set of tactile models can support multi-step movement learning, even without dynamic cues.

Overall, 3D tactile models not only convey individual pose shapes but also create new opportunities for embodied, hands-on learning of movement. By combining spatial clarity with physical manipulability, they provide a tangible foundation for understanding movement transitions and body coordination, addressing gaps left by existing non-visual instructional methods.

\subsection{Toward Multimodal and Dialog-Based Support}

The 3D tactile models effectively supported understanding of individual poses; however, some participants in User Study~2 reported challenges when learning sequences. As the number of poses increased, the memory load also grew, and several participants noted difficulty remembering the correct order or how to move between poses. \ken{This issue aligns with prior research by Holloway et al.~\cite{Holloway2022}, which noted that when representing dynamic motion through tactile displays, the number and granularity of frames must be adjusted based on the complexity of the movement being conveyed.} Static models were sufficient for the four-pose sequences used in this study, but these comments suggest that longer or more complex routines may require additional forms of support to ease cognitive demands and clarify movement transitions.

To address these limitations, participants proposed integrating tactile exploration with complementary modalities. One frequently mentioned option was integrating step-by-step audio guidance with 3D tactile models. \ken{Prior work has also suggested that combining tactile and auditory information can support understanding of dynamic motion while requiring careful balance to avoid cognitive overload~\cite{Mackowski2023}. Building on this approach, a learner could first explore the overall form of a pose through touch, then trigger an audio segment describing how to shift weight, rotate joints, or progress smoothly between steps. This staged method would preserve the strengths of tactile comprehension while enhancing clarity and confidence in performing the full sequence.}

Several participants also envisioned systems offering real-time confirmation of body posture. Such functionality could be achieved using wearable motion sensors or computer vision to detect deviations from the intended pose and provide immediate corrective audio or vibrotactile feedback. \ken{Similar concepts have been explored in vision-based systems such as by Rector et al.~\cite{RectorYoga2014, RectorYoga2013, Rector2017, Rector2018} and Dias et al.~\cite{Dias2019}, which use Kinect-based tracking to analyze users’ poses and deliver feedback to improve alignment and movement accuracy. Incorporating comparable sensing techniques could enable learners to self correct without external supervision while maintaining the benefits of tactile exploration.}

\ken{Finally, dialog-based interaction could further personalize learning. Through simple voice commands, learners could repeat transitions, slow playback, or check limb orientation. Reinders et al.~\cite{Reinders2025} found that multimodal Interactive 3D Models (I3Ms) combining tactile, auditory, and conversational feedback make exploration more embodied and engaging for BLV learners. However, prior studies have noted drawbacks such as higher cognitive load, divided attention, and timing mismatches when multiple channels are processed simultaneously~\cite{Mackowski2023,Mackowski2022,Morrison2005,Paas_Sweller_2014}. Future systems should balance informational richness and pacing to prevent cognitive overload.}

Taken together, these suggestions outline a pathway toward multimodal, interactive systems that combine tactile, auditory, and responsive feedback channels. Such integration could reduce memory burden, improve accuracy in sequential tasks, and make movement learning more scalable for people with visual impairments.

\subsection{Physical Design Considerations for Identifiability and Orientation}
All participants found the tactile reference markers on the model bases useful for understanding foot placement and forward–backward movement. These markers helped clarify body direction and supported interpretation of front-back orientation during exploration.

While participants were generally able to explore the models without confusion, future designs may benefit from added cues to support orientation and identification. As sequences grow longer or involve rotational movements, it may become harder to distinguish between poses or to determine the model’s front side. \ken{De Silva et al.~\cite{de_silva_sensing_2025} noted that unclear directional cues can cause confusion for blind learners and that spatial orientation is a crucial factor supporting movement understanding. In our study, participants similarly suggested adding tactile numbers on the model bases and fixed directional markers to indicate forward facing orientation, which could help maintain consistency during exploration.}

One strength of our approach is the use of automatic posture estimation (Humans in 4D) to generate 3D tactile models. This enabled efficient extraction of key body configurations from motion data, reducing the need to design each pose manually and preserving natural human movement. However, several manual steps were still required, such as selecting keyframes, refining meshes, and adding bases for tactile stability. \ken{Recent work has explored automated keyframe extraction from dance videos using both audio and visual cues to identify rhythmically and semantically meaningful motion segments~\cite{tokida2024, endo2021}.} Future work should focus on automating this pipeline—streamlining frame selection and 3D model generation—and incorporating additional cues like audio or rhythm to better support learning of multi-step movements.

\subsection{Scope and Generalizability of Findings}

\ken{Although this study yielded valuable insights, its scope was limited to 10 participants and one co-designer. While this small scale enabled rich qualitative feedback and close iterative refinement, it may also reflect individual preferences that influenced design outcomes. Future research should include participants and co-designers with more diverse backgrounds and experiences in physical activity. Cultural familiarity with exercises such as yoga or calisthenics may also affect how easily movements are learned, underscoring the need for cross-cultural validation.}

In addition, the study focused on a limited set of movements, primarily static yoga poses and a simple calisthenics sequence. To understand the broader applicability of tactile pose models, future work should evaluate their effectiveness across a wider variety of movement types, including more dynamic, asymmetrical, or unfamiliar actions. Long-term assessment of learning retention and real-world usability will also be important for determining the practical impact of such systems.

\ken{Beyond experimental settings, the proposed 3D tactile models are primarily envisioned for independent home-based learning by people who are blind or have low vision, enabling them to practice physical exercises at their own pace. Future work will further explore these real world use contexts to understand how tactile models can best fit into daily learning routines. In addition, future iterations could investigate physically manipulable or actuated models that convey transitions between poses, allowing learners to perceive dynamic movement sequences through tangible feedback.}

%% file: SECTIONS/8_Conclusion.tex
\section{CONCLUSION}

This study presented a 3D tactile model based approach that enables blind individuals to learn body movements and postures through direct, hands on exploration, addressing limitations of existing audio and 2D tactile methods in conveying full body spatial relationships. Through two user studies with ten participants, we evaluated its effectiveness for both static yoga poses and sequential calisthenic movements. Results showed that 3D tactile models provided clearer spatial and mental representations, increased participants’ confidence and motivation, and offered a reproducible, scalable workflow for producing tactile learning materials using pose estimation technology.

\ken{By demonstrating the feasibility and benefits of this embodied learning approach, our work lays the groundwork for future systems that combine tactile exploration with complementary modalities such as audio guidance or real-time feedback. Although this study was limited in scale and scope, expanding evaluations to larger and more diverse participant groups and varied activity contexts will further validate and refine these methods. Future research should develop multimodal systems that allow BLV users to learn and practice movements independently, integrating tactile and auditory feedback for self-paced, accessible learning. Together, these directions point toward more personalized and embodied approaches that advance inclusive design in physical education.}